\documentclass[twocolumn,superscriptaddress,floatfix,prb,aps,showpacs]{revtex4}
\usepackage{graphicx}
\begin{document}
\bibliographystyle{apsrev}
\def\sa{\section}
\def\sb{\subsection}
\def\sc{\subsubsection}

\def\ind{\ \ \ \ }

\def\der{\partial}

\def\be{\begin{equation}}
\def\ee{\end{equation}}

\def\bea{\begin{eqnarray}}
\def\eea{\end{eqnarray}}

\def\ba{\begin{array}}
\def\ea{\end{array}}

\def\nn{\nonumber}

\def\ben{\begin{enumerate}}
\def\een{\end{enumerate}}

\def\fn{\footnote}

\def\rd{\partial}
\def\rot{\nabla\times}

\def\r{\right}
\def\l{\left}

\def\gt{\rightarrow}
\def\cf{\leftarrow}
\def\bw{\leftrightarrow}

\def\ra{\rangle}
\def\la{\langle} 
\def\bla{\big\langle}
\def\bra{\big\rangle}
\def\Bla{\Big\langle}
\def\Bra{\Big\rangle}
\def\bbla{\bigg\langle}
\def\bbra{\bigg\rangle}

\def\ddt{{d\over dt}}

\def\rdt{{\rd\over\rd t}}
\def\rdx{{\rd\over\rd x}}

\def\bb{\begin{thebibliography}{99}}
\def\eb{\end{thebibliography}}
\def\bit{\bibitem}

\def\bc{\begin{center}}
\def\ec{\end{center}}

\title{Characterization of 2D fermionic insulating states}
\author{Ryuichi Shindou}
\affiliation{Department of Physics, University of Tokyo,
Bunkyo-ku, Tokyo 113-8656, Japan.}
\affiliation{Department of Physics, University of California, 
Santa Barbara, California 93106, USA}
\author{Ken-Ichiro Imura}
\affiliation{Condensed Matter Theory Laboratory,
RIKEN (Wako), Hirosawa 2-1, Wako, 351-0198, Japan.}
\author{Masao Ogata}
\affiliation{Department of Physics, University of Tokyo,
Bunkyo-ku, Tokyo 113-8656, Japan.}
\date{\today}
\begin{abstract}
Inspired by the duality picture between superconductivity (SC) and insulator
in two spatial dimension (2D), we conjecture 
that the order parameter, suitable for characterizing 2D fermionic 
insulating state, is the {\it disorder operator},
usually known in the context of statistical transformation. 
Namely, the change of the phase of the disorder operator 
along a closed loop measures the particle density accommodating 
inside this loop. Thus, identifying this (doped) particle density 
with the dual counterpart of the magnetic induction in 2D SC, 
we can naturally introduce the disorder operator as the 
dual order parameter of 2D insulators. 
The disorder operator has a branch cut emitting from 
this ``vortex'' to the single infinitely far point.   
To test this conjecture against an arbitrary 2D lattice models,
we have chosen this branch cut to be compatible with the 
periodic boundary condition and obtain a general form of 
{\sl its expectation value 
for non-interacting metal/insulator wavefunction, including gapped 
mean-field order wavefunction}. Based on this expression, we observed 
analytically that it indeed vanishes for a 
wide class of band metals in the thermodynamic limit. 
On the other hand, it takes a finite value in insulating states, 
which is quantified by the {\it localization length} or 
the real-valued gauge invariant 2-from dubbed as the 
{\it quantum metric tensor}.  
When successively applied along a closed loop, our 
disorder operator plays role of {\sl twisting the boundary 
condition of a periodic system}. 
We argue this point, by highlighting the 
Aharonov-Bohm phase associated with this non-local operator. 
\end{abstract}

\maketitle

\section{Introduction}

The high-$T_{\rm c}$ cuprates and the colossal magnetoresistive manganites, 
the two main realms in condensed matter physics,
are both doped correlated insulators.
As type II superconductors accommodate magnetic flux
by allowing a spatially inhomogeneous distribution of the superconducting 
order parameter, 
such fermionic insulating states do not necessarily remain spatially
uniform {\sl against doping.}
Indeed recent STM experiments have observed that
a hole rich region in a doped Mott insulator such as Bi-2212
forms a nanoscale granular island on top of 
the insulating background~\cite{lang}.
A similar kind of nanoscale electronic inhomogeneity 
was also found in Ca$_{2-x}$Na$_x$CuO$_2$Cl$_2$, 
where doped holes form a checkerboard pattern~\cite{hana}.
On the other hand, extensive Lorentz optical microscopy measurements 
on doped (e.g., mixed valence) manganites have 
revealed that charge ordered and ferromagnetic metallic patches 
are phase-separated in such relatively large length scales as a few
micrometer~\cite{uehara}.
The question naturally arises as follows :
Is there some universal classification of insulators in 
terms of their different types of behavior against doping? 
The other way around, if it exists, 
what kind of {\sl microscopic ingredients}  
would determine these differences? 

%
%
These experimental questions/observations as well as almost 
all the other experimental observables 
are directly accessible through the ``local electronic polarization'' 
$\vec{P}(\vec{r})$, which acquires finite {\it mass} only in  dielectrics. 
Contrary to other observables, 
however, the local electronic polarizations   
cannot be uniquely determined by those 
vector fields which are 
divergence-free. Namely, using arbitrary {\it analytic} scalar 
functions $\theta(\vec{r})$, we can introduce 
those vector fields which have nothing to do with the local 
charge density $\rho(\vec{r})\equiv  
\vec{\nabla}\cdot\vec{P}(\vec{r})$;    
\begin{eqnarray}
\vec{P}'(\vec{r})&=&\vec{P}(\vec{r}) + 
\times\vec{\nabla}\theta(\vec{r}), \nonumber \\
\vec{\nabla}\cdot \vec{P}'(\vec{r})&=& 
\vec{\nabla} \cdot \vec{P}(\vec{r}). \label{polariz} 
\end{eqnarray}  
Due to this arbitrariness,  constructing 
effective theories of dielectrics in such a way that {\it the 
local polarization becomes explicit} has been one of the  
long-standing issues in condensed matter theories community, 
while widely demanded from experimental sides 
on a general ground.  

The central idea of the duality picture 
is to regard this local electronic polarization as 
the ``gauge field'' {\it which is intrinsic in matters} 
and to treat its divergence-free part as the 
unphysical ``gauge'' degree of freedom.   
\cite{fisher,balents1,balents2,tesanovic}
To be more specific, 
we suppose that there exist  
{\it order parameters}, implicit in a microscopic model,
 which are coupled with the local electronic  
polarization such that its condensation makes this ``gauge 
fields'' massive via the Higgs mechanism 
\cite{balents1,balents2,tesanovic}.  Accordingly,
the ``gauge'' in eq.(\ref{polariz}) and that of this  
order parameter $\eta(\vec{r})$ are specified  
{\it in a set}, as in superconductors;    
\begin{eqnarray}
\eta'(\vec{r})=\eta(\vec{r})e^{i\theta(\vec{r})}. 
\end{eqnarray}   
where the vortex-free scalar function $\theta(\vec{r})$ 
should be identical to that in Eq.~(\ref{polariz}).   

In this paper, observing this duality picture, we will introduce 
this complex-valued order parameter $\eta(\vec{r})$ 
explicitly {\it in terms of original fermion (electron) operators}. Then  
we calculate the expectation value of this order parameter 
with respect to some simple wavefunctions, so that we 
can argue that this order parameter has indeed a finite 
amplitude, i.e. {\it condensed}, only in fermionic insulators.     

This paper is organized as follows :
In Sec.~II, we look further into the duality between 
superconductivity and insulator 
only to arrive at the conjecture that the appropriate order
parameter for a 2D fermionic insulating state is 
the {\sl disorder operator (DOP)}  
defined as eqs.~(\ref{plane},\ref{torus}).
We then give a general expression to the expectation 
value of the DOP. 
In Sec.~III, we will see that this expectation value 
indeed vanishes for band metals having 
various kinds of Fermi surface (F.S.) in the thermodynamic 
limit. In Sec.~IV, we will 
consider the opposite limit, i.e., the case of 
a band insulator/gapped mean-field order state 
close to the atomic limit. Thereby, we observe that 
the expectation value of the DOP 
is in turn characterized by the localization length 
and thus remain finite. Then these two observations, i.e. 
those in sec. III and in sec. IV,  
lead us to extrapolate the behaviour of the DOP 
in general insulating states.   
Sec.~V is devoted to the arguments on the relation between the 
DOP and ``twisting boundary condition'', the latter known to give a 
definite criterion for insulating states in arbitrary 
spatial dimension~\cite{kohn, kudinov, ivo, scalapino}. 
Sec.~VI contains not only the brief summary (see Table. II) but 
also discussions on the behaviour of our DOP 
in the off-diagonal long ranged 
ordered states. We also mentioned there about  
the possible microscopic candidates of 
{\it the counterpart of the magnetic penetration 
depth/coherence length} (see Table. I) and the future possible progress. 
Some details of the calculation are  
left to the appendix.

\section{The disorder operator and the insulating order parameter}

\subsection{On duality between superconductivity and insulator}

Let us denote the usual superconducting order parameter as
$\Delta(\vec{r})=|\Delta(\vec{r})| e^{i\phi (\vec{r})}$. In this 
paper, we consider only  
 two spatial dimension (2D), i.e. $\vec{r} = (x,y)$.
If the magnetic penetration depth $\lambda$ is large enough 
compared with the coherence length $\xi$
(if $\kappa \equiv \lambda/\xi>1/\sqrt{2}$ in the conventional 
Ginzburg-Landau theory~\cite{deG}), 
then the system shows a type II superconducting behavior,
accommodating magnetic flux $b(\vec{r})$ 
 pinned to {\sl vortices} in the system.
The amplitude of order parameter $|\Delta(\vec{r})|$ is
spatially not uniform and can have zeros.
Then its phase $\phi (\vec{r})$ acquires an ambiguity of integer
multiples of $2\pi$ around its zeros.
The holonomy of this phase around such a vortex 
is proportional to the number of magnetic flux pinned
to this vortex, counted in units of the flux quantum
$\Phi_0\equiv \hbar c/(2e)$. 
Let a flux penetrate a specific area $S$. Then, we have,   
\bea
\Phi_0\oint_{\der S} \nabla \phi (\vec{r}) \cdot d\vec{l}=  
\int_{\vec{r} \in S} b(\vec{r})\ d^2 r,
\eea
where $\nabla \equiv (\der/\der x, \der/\der y)$ and $\der S$
is the boundary of this area $S$. 
We now invoke this relation to identify the {\sl dual quantity} of 
$\Delta(\vec{r})$. 

\begin{table}[htbp]
\begin{center} 
\begin{tabular}{c|c}
\hline
2D electronic insulator  &  2D superconductor \\ 
\hline \hline
doping & applied magnetic field \\
doped particle density: & 
magnetic induction:  \\
$\delta\rho(\vec{r})=\rho(\vec{r})-\bar{\rho}$ & $b(\vec{r})$ \\ \hline
``type I insulator (?)'' & type I superconductor \\
``type II insulator (?)'' & type II superconductor \\  
 ? & penetration depth \\
 ? & coherence length \\ \hline \hline 
disorder operator: & order parameter: \\
$\eta(\vec{r})=|\eta(\vec{r})|e^{i\theta(\vec{r})}$ & 
$\Delta(\vec{r})=|\Delta(\vec{r})|e^{i\phi(\vec{r})}$ \\ \hline   
$a_{\mu}(\vec{r})=\epsilon_{\mu\nu}P_{\nu}(\vec{r})
=\nabla_{\mu}\theta(\vec{r})$ &
$A_{\mu}(\vec{r})=\nabla_{\mu}\phi  (\vec{r})$ \\
$\delta\rho(\vec{r})=\nabla\cdot\vec{P} = \nabla\times a$ & 
$\delta b(\vec{r})=\nabla\times A$ \\
\hline
\end{tabular} 
\end{center}
\caption{The duality between 2D insulator and superconductivity}
\end{table} 

In the duality relation between 2D superconductivity and 2D insulator, 
applying magnetic field in superconductors  
corresponds to doping in insulators~\cite{lee1} (Table. I).  
Thus the doped particle density 
$\delta\rho(\vec{r})\equiv\rho(\vec{r})-\bar{\rho}$ 
corresponds to magnetic flux in superconductors. 
Physically speaking, $\delta\rho(\vec{r})$ is obtained from a 
{\sl local electronic polarization} $\vec{P}(r)$;  
$\delta \rho(\vec{r})=\vec{\nabla}\cdot\vec{P}(r)$. Then
introducing the ``gauge field'' $\vec{a}(\vec{r})$ 
such that ${a}_{\mu}(\vec{r})=\epsilon_{\mu\nu}P_{\nu}(\vec{r})$, 
we have only to find out a scalar function whose gradient is 
this ``gauge field'', i.e. 
$\partial_{\mu}\theta(\vec{r})=a_{\mu}(\vec{r})$. 
Namely, we can identify this scalar function 
$\theta (\vec{r})$ as the phase part of the 
dual order parameter, i.e. counterpart of $\phi (\vec{r})$.
We thus reach the conjecture that the  
{\sl disorder operator (DOP)} 
$\tilde{\eta} (\vec{r})$~\cite{frad} 
should play role of an insulating order parameter; 
\be
\tilde{\eta} (z)\equiv {\rm exp}\Big[ \int_{z^{\prime}\ne z}d^2 z^\prime 
\log (z^{\prime}-z)\l\{\rho(z^{\prime})-\bar{\rho}\r\} \Big].
\label{plane} 
\ee
Here we have introduced a complex variable $z=x+iy$, and
$d^2 z$ should be understood to be $d^2 z=d\bar{z}dz/(2i)$.
Note that $\tilde{\eta}(\vec{r})$ is defined on 
a 2D infinite plane {\sl without any boundary condition}.
Using this explicit form, one can readily verify that 
eq.~(\ref{plane}) indeed satisfies our leading principle,
\be
{1\over 2\pi}\oint_{\partial S} \nabla{\rm arg}\ \tilde{\eta}(\vec{r})\cdot 
d\vec{l}=
\int_{\vec{r}\in S} \l\{ \rho (\vec{r})-\bar{\rho} \r\} d^2 \vec{r}.
\label{vortex}
\ee
This states that the winding number associated with the dual 
order parameter $\tilde{\eta}(\vec{r})$ is identical to the
doped particle number inhabiting within $S$.

\subsection{Disorder operator for a periodic lattice}

Table. I shows the correspondence between 2D insulator and
superconductivity.
In the middle two empty seats are reserved for the {\sl unknown} 
counterparts of the magnetic penetration depth and the
coherence length in 2D insulator.
If we push forward with this duality relation, 
our eventual goal might be to construct a Ginzburg-Landau theory for 
2D (doped) insulating state in terms of this non-local operator, 
which unambiguously lets us complete this 
table {\sl quantitatively}.   
As a first step toward this direction, we will demonstrate 
in the following sections that the expectation value of 
the DOP  
\ben
\item
indeed vanishes in various types of 2D band metals, i.e., 
the DOP is {\sl not condensed} (see Sec.~III), whereas
\item
it remains finite in 2D band insulators, i.e., 
the DOP is {\sl condensed} (see Sec.~IV). 
\een

If one tries to make sure these statements,  
one might immediately notice that the DOP defined 
in eq.~(\ref{plane}) in itself is incompatible with the {\sl periodic 
boundary condition (PBC)} which we usually presume.
The branch cut of logarithm in eq.~(\ref{plane}) 
has only a single end point, while, mathematically speaking, 
it is impossible to embed such a branch 
cut into a torus.   

\begin{figure}
\begin{center}
\includegraphics[width=0.4\textwidth]{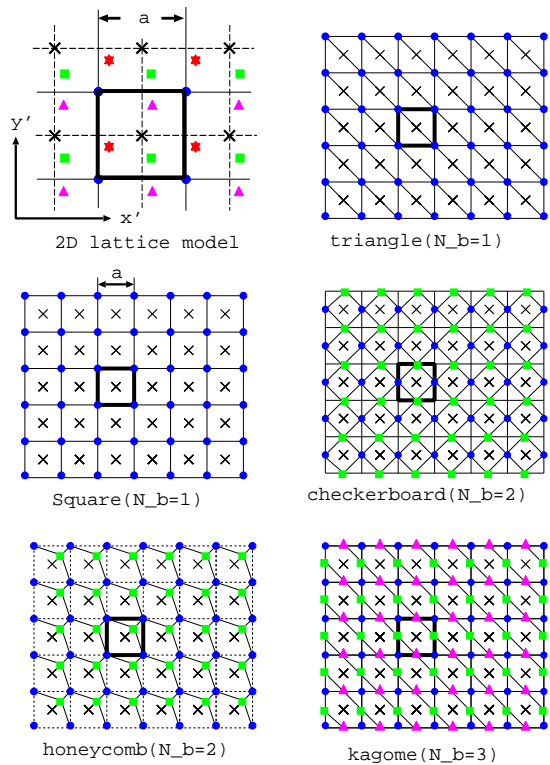}
\end{center}
\caption{ (color online) 2D lattice models. $(x,y)$ and $(x',y')$  
specify a lattice point (filled circle mark).  
A unit cell is depicted as bold line in every figure and we 
use different colors (shapes)  in order to distinguish $N_b$ number 
of inequivalent lattice points within it. Crossed mark represents 
a dual lattice site on which $(x+x_0,y+y_0)$ is located.     
(From top to down): Triangle and Square lattice 
with $N_b=1$, honeycomb and checkerboard lattice 
with $N_b=2$ and kagome lattice with $N_b=3$. 
Note that $(x+x_0,y+y_0)$ (crossed mark) never 
coincides with the original lattice sites  
on which a fermion operator is defined. }
\label{1}
\end{figure}

\begin{figure}
\includegraphics[width=0.45\textwidth]{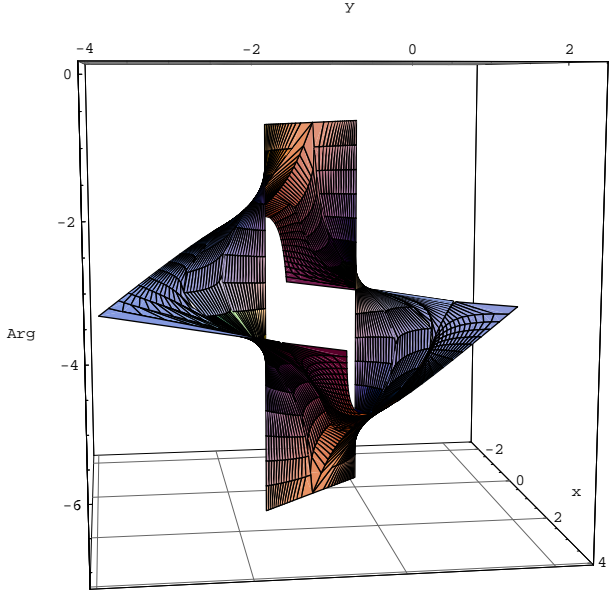}
\caption{ (color online) 
${\rm Arg}{\rm log}[-i(e^{ix}-1)+(e^{iy}-1)]$ as a function of $x$ and $y$, where 
we only show the principle branch which ranges from 
$-\frac{\pi}{4}$ to $-\frac{9\pi}{4}$ in this picture 
and those below. }
\label{2}
\end{figure}

\begin{figure}
\begin{center}
\includegraphics[width=0.25\textwidth]{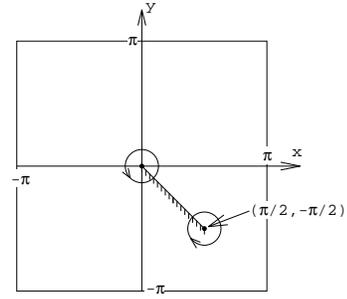}
\end{center}
\caption{
: A branch cut runs from the vortex $((x,y)=(0,0))$ to the anti-vortex 
$((x,y) = (\frac{\pi}{2},-\frac{\pi}{2}))$. 
${\rm Arg}{\rm log}[-i(e^{ix}-1)+(e^{iy}-1)]$ takes 
$-\frac{\pi}{4} - $ in the shaded side of 
this branch cut, while takes $- \frac{9\pi}{4} +$ in the other side. }
\label{3}
\end{figure}

A possible and the most plausible way out would be to
replace $z'-z$ in this logarithm 
by a doubly periodic function $\zeta(x'-x,y'-y)$, i.e. 
$\zeta(x, y + L)\equiv \zeta (x+L,y)\equiv \zeta(x,y)$,  
{\sl which reproduce $z'-z$ when $|z'-z|\ll L$}, 
\bea
\lim_{|z'-z|/L \gt 0}
\zeta(x'-x,y'-y)\simeq z'-z. \label{requ}
\eea
Here $L$ denotes the linear dimension of a system size. 
As the simplest function satisfying this requirement, 
we consider in this paper the following function,
\bea
&&\zeta(x'-x,y'-y)\equiv \nn \\
&&\ \ -i\big(e^{i{2\pi\over L}(x'-x-x_0)}-1\big) 
+\big(e^{i{2\pi\over L}(y'-y-y_0)}-1\big).  \label{zetadef}
\eea
Apart from the continuum variables $z$ and $z'$ used 
in eq.~(\ref{plane}), $\vec{r}\equiv (x,y)$ and $\vec{r}'\equiv (x',y')$ 
in eq.~(\ref{zetadef}) are defined to take a {\sl discrete value},  
specifying lattice points in a particular lattice model. 
$(x_0,y_0)$ was then introduced, so as to  
{\sl avoid the logarithmic singularity when $\log\zeta$ 
are summed with respect to $\vec{r}'$ over all   
lattice points (see eq.~(\ref{torus}))}.  
In a simple square lattice with 
its lattice constant ``a'',  
we will take $(x_0,y_0)\equiv(\frac{\rm a}{2},\frac{\rm a}{2})$
(see Fig.~\ref{1}). 
This function, i.e. eq.~(\ref{zetadef}),  in fact  
satisfies the requirement eq.~(\ref{requ}). Namely, 
when $|z'-z| \ll L$, it actually reads, 
\[
\zeta(x'-x,y'-y)\simeq {2\pi\over L}\{(x'-x-x_0)+i(y'-y-y_0)\}. 
\]
It is also a doubly periodic function. 
As a result, when seen as a continuous function of $(x',y')$, 
$\zeta(x'-x,y'-y)$ has two zeros: 
\bea
(x', y') = (x+x_0,y+y_0),\ \
\big(x+x_0+{L\over 4},y+y_0-{L\over4} \big). \label{zeros}
\eea
The first zero corresponds to the original vortex introduced 
in eq.~(\ref{plane}). The 
latter one turns out to be the {\sl antivortex}, which was 
supposed to be located at the single infinitely far point in 
eq.~(\ref{plane}). 
To be more specific, 
when $(x', y')$ moves around the former (latter) zero point 
anti-clockwise, then $\log\zeta$ picks up a phase $2\pi$ ($-2\pi$), 
as shown in Fig.~\ref{2}. Accordingly, $\log\zeta$ has a branch cut 
running from the vortex to the antivortex as in Fig.~\ref{3}.
In general, the singularities must appear in pairs in a 
system obeying PBC. 
In order to study the DOP 
explicitly in non-interacting states and 
mean-field ordered states, it is necessary to {\sl write down a  
ground state wavefunction explicitly}. 
Without PBC, we would not be able to do this. 
We thus conjecture in the remainder of the paper that
the following operator with $\zeta$ defined in eq.~(\ref{zetadef}) 
is {\sl the appropriate form of the disorder operator (DOP) 
compatible with PBC}, 
\be
\eta(\vec{r})\equiv\exp\big[
\sum_{x',y'}\log \zeta (x'-x, y'-y)
\{\rho(\vec{r'})-\bar{\rho}\}
\big]. 
\label{torus}
\ee

Notice that the $z'$-integral in eq.~(\ref{plane}) avoids $z'=z$, 
while the summation w.r.t. $\vec{r}'\equiv (x',y')$ 
in eq.~(\ref{torus}) is taken over entire lattice 
points {\sl without any restriction}.   
As we mentioned above in the square lattice case 
with its lattice constant ``a'', this simplification 
becomes possible, just because we have 
introduced $(x_0,y_0)=(\frac{\rm a}{2},\frac{\rm a}{2})$ 
so that $\log\zeta(n_x{\rm a},n_y{\rm a})$ is always 
finite for arbitrary integer $n_x$ and $n_y$. 
All 2D lattice models, however,  
can be also regarded  as having a square unit cell (See Fig.~\ref{1}).  
Then, we can always choose 
$(x_0,y_0)$ appropriately, 
{\sl such that the vortex and antivortex introduced in eq.~(\ref{zeros})
never coincide with any lattice points on which $\rho(\vec{r}')$ is defined 
(see some examples shown in Fig.~\ref{1}).}  
Provided that 
$\log \zeta (\vec{r}'-\vec{r})$ in eq.~(\ref{torus}) takes 
finite value for an arbitrary lattice point $\vec{r}'$, then the 
following arguments do not depend seriously on the 
specific choice of $(x_0,y_0)$.

Before giving the general expression to 
the expectation value of our DOP defined in eqs.~(\ref{zetadef},\ref{torus}), 
let us fix our ideogram of this paper now. 
Firstly, we always refer a lattice constant of a 
unit cell (square box depicted by bold line in Fig.~\ref{1}) 
to ``$\rm a$''. We take a number of this unit 
cell in an entire system to be 
$N \times N$ and thus the linear dimension of a system size 
is given by $L=N\rm a$. We call as $N_b$, the total 
number of Bloch bands/inequivalent sites 
within each unit cell, while total  
number of electrons which a ground state w.f.  has is denoted by 
$N_e$. Then the average particle number {\sl per site} $\bar{\rho}$ 
introduced in eq.~(\ref{torus}) is given as follows, 
\bea
\bar{\rho} \equiv \frac{N_{e}}{N_b\cdot N^2}. 
\eea  


\subsection{The general expression for expectation values of the disorder 
parameter}
For a non-interacting system and arbitrary mean-field ordered state  
with $N_e$ fermions, a ground wavefunction  
is obtained just by filling the lowest $N_e$ one-body states 
which we name simply as $\alpha=1,2,\cdots,N_e$, 
\bea
|{\rm GS}\ra=c_{N_e}^\dagger c_{N_e-1}^\dagger\cdots
c_{2}^\dagger c_{1}^\dagger|0\ra. \label{groundstate}
\eea
Here $|0\ra$ denotes the fermion-vacuum state. 
To evaluate our DOP for such a ground state, 
let us apply our DOP onto a creation or
an annihilation operator.  We define the matrix elements 
$\zeta [\alpha|\beta]$ as 
\be
\eta(\vec{r})c^{\dagger}_\alpha=
\sum_\beta 
c_\beta^\dagger \zeta [\beta|\alpha] \eta(\vec{r}),
\label{zeta1ab}
\ee
where $\alpha$ and $\beta$ specify a one-body electronic state.
For sake of simplicity, we will make the $\vec{r}$ 
dependence of $\zeta [\alpha|\beta]$ implicit from now on. 
Since $\eta(\vec{r})$ is nonlocal,
it does not commute even with creation and annihilation
operators at different positions ($\vec{r'}$).  From the 
expression of $\eta(\vec{r})$ in eq.~(\ref{torus}), 
we can readily see this,  
\bea
\eta(\vec{r}) c^\dagger (\vec{r'})=
\zeta (x'-x,y'-y)
c^{\dagger}(\vec{r'})\eta(\vec{r}). \label{real}
\eea
On comparing this with eq.~(\ref{zeta1ab}), we also see 
that $\zeta[\alpha|\beta]$ is {\sl diagonal in this real-space 
representation}, 
\be
\zeta [\vec{r'_1}|\vec{r'_2}]
=\delta(\vec{r'_1}-\vec{r'_2})\zeta (x'_1-x, y'_1-y).
\label{zeta2ab}
\ee
In the momentum-space representation, however, 
$\alpha$ and $\beta$ is 
specified by the crystal momentum $\vec{k}$ 
and the band index $n=1,2,\cdots,N_b$.  
The creation operator of such a Bloch state 
is defined as follows,
\[
c^{\dagger}_{n,\vec{k}}= 
\frac{1}{N}\sum_{\vec{r}}
e^{-i\vec{k}\cdot\vec{r}}u_{n,\vec{k}}(\vec{r})
c^{\dagger}_{\vec{r}},
\]
where 
$u_{n,\vec{k}}(\vec{r})$ is the periodic part of the Bloch wavefunction.  
Then, substituting the above equation into 
eq.~(\ref{real}), we readily obtain the explicit matrix elements 
for $\zeta [\alpha|\beta]$ in this momentum representation, 
\bea
&&\zeta [n,\vec{k}|n',\vec{k'}]  
=(i-1)\delta_{n,n'}\delta_{\vec{k},\vec{k'}}
\nn \\
&&-i e^{-i{2\pi\over L}(x+x_0)}
\Bla u_{n,(k_x,k_y)}\Big|u_{n',(k_x+{2\pi\over L},k_y)}\Bra 
\delta_{k_x,k'_x-{2\pi\over L}}\delta_{k_y,k'_y}
\nn \\
&&+e^{-i{2\pi\over L}(y+y_0)}
\Bla u_{n,(k_x,k_y)}\Big|u_{n',(k_x,k_y+{2\pi\over L})}\Bra 
\delta_{k_x,k'_x}\delta_{k_y,k'_y-{2\pi\over L}}. \nn \\
\label{zeta3ab}
\eea
where the inner product between $|u_{n,\vec{k}}\ra$ 
represents an integral over the unit cell.  
These inner products are nothing but the gauge connections 
in $\vec{k}$ space. Note that, in this momentum 
representation, $\zeta [\alpha|\beta]$  
is no longer diagonal. Instead, it takes a matrix form 
representing a {\sl $\vec{k}$-derivative} or 
{\sl $\vec{k}$-covariant derivative}.  
The above relations defined in eqs.~(\ref{zeta1ab})-(\ref{zeta3ab})
play a fundamental role in the reminder of the present paper.

\subsubsection{The determinant formulae}

Using the ground state wavefunction defined in eq.~(\ref{groundstate}), 
let us formulate a general expression for its expectation 
value of the DOP,  
\be
\la \eta(\vec{r})\ra=\la{\rm GS}|\eta(\vec{r})|{\rm GS}\ra.
\label{GS}
\ee
Noticing that for an arbitrary permutation of $N_e$ quantum numbers,
\bea
&&{\rm sgn}(P) = \nn \\
&&\la 0|c_{1}c_{2}\cdots c_{N_e-1}c_{N_e}
c_{P(N_e)}^\dagger c_{P(N_e-1)}^\dagger\cdots
c_{P(2)}^\dagger c_{P(1)}^\dagger|0\ra, \nn
\eea
one can readily verify,
\bea
&&\la \eta(\vec{r})\ra=\la 0|\eta(\vec{r})|0\ra \sum_{P}{\rm sgn}(P) \nn \\
&&\times \zeta[P(N_e)|N_e]\cdots\cdot\zeta[P(2)|2]\cdot\zeta[P(1)|1],
\label{det1}
\eea 
where 
$\la 0|\eta(\vec{r})|0\ra$
is a contribution from the uniform background $\bar{\rho}$.
Leaving further analysis on $\la 0|\eta(\vec{r})|0\ra$ 
to the following subsection, let us for the moment consider 
eq.~(\ref{det1}).  
The summation over $P$ should be taken for all the possible 
permutations of $N_e$ {\sl occupied states}. 
Here, let us use eq.~(\ref{zeta3ab}) as $\zeta[\alpha|\beta]$, 
bearing in mind a band metal/insulator. 
Then, in order to interpret the summation over $P$ 
as a determinant, we will arrange or rearrange every row and column 
in such a way that {\it all the matrix elements $\zeta [\alpha|\beta]$ with 
$\alpha$ and $\beta$ being occupied Bloch states 
should be accommodated in the upper-left block of $\zeta$}. 
Let us call this $N_e\times N_e$ submatrix as $\zeta'$, i.e.,
\[
\zeta=\l(\ba{ccc|c}
 &  & &   \\
 &\zeta'& &* \\ 
 &  & &    \\\hline
 &* & &* \\
\ea\r).
\]
Then one can reinterpret eq.~(\ref{det1}) as
\be
\la \eta(\vec{r})\ra=\la 0|\eta(\vec{r})|0\ra\det\zeta'.
\label{det2}
\ee
Note that the final formula (\ref{det2}) itself is 
independent of the representation.
In fact, when $|\rm GS\ra$ were given by a Slater determinant composed by 
{\sl atomic orbitals totally localized in the real space}, then we would use
 eq.~(\ref{zeta2ab}) instead of eq.~(\ref{zeta3ab}) as 
 $\zeta$ and thus $\zeta'$.  
However, a non-interacting band metal/insulator 
wavefunction (w.f.), including gapped mean-field w.f., is usually 
composed by extended Bloch w.f.. Thus, we will  
evaluate eq.~(\ref{det2}), using mainly its 
momentum representation.  

\subsubsection{Contribution from the uniform background}

Before ending this section, let us estimate the contribution 
from the uniform background, i.e. $\la 0|\eta(\vec{r})|0\ra$.
From the definition of eq.~(\ref{torus}), it reads, 
\be
\log\la 0|\eta(\vec{r})|0\ra=-\bar{\rho}
\sum_{x',y'}\log\zeta (x'-x,y'-y).
\label{eta0zeta}
\ee
By using the real-space representation of the matrix $\zeta$, 
i.e. eq.~(\ref{zeta2ab}), it could be rewritten as 
\[
\log \la 0|\eta(\vec{r})|0\ra=-\bar{\rho}\log\det \zeta.
\]
Recall that $\bar{\rho}$ is the average particle number {\sl per site}; 
$\bar{\rho}\equiv \frac{N_{e}}{N_{b}N^2}$. 
Thus, in terms of the filling fraction {\sl per site} $\nu\equiv\bar{\rho}$, 
$\la 0|\eta(\vec{r})|0\ra$ has the following compact form, 
\bea
\la 0|\eta(\vec{r})|0\ra=\frac{1}{(\det\zeta)^{\nu}}. \label{remark}
\eea
We will use this expression in section.~IV.  

Apart from its compact re-expression, eq.~(\ref{eta0zeta}) itself 
can be directly evaluated. Notice that we have only to 
estimate the {\sl amplitude} of
$\la\eta (\vec{r})\ra$, so as to check whether it (does not) vanishes 
in a metal (insulator) or not. 
Thus we will focus only on the real part of 
$\log \la 0|\eta(\vec{r})|0\ra$. 
At the leading order in the thermodynamic limit, 
i.e., at the order of ${\cal O}(L^2)$, the summation with respect to (w.r.t.)  
$\vec{r'}=(x',y')$ can be replaced by an integral 
(see appendix.~A), 
\bea
&&\log \la 0|\eta(\vec{r})|0\ra= -\frac{N_e}{4\pi^2} \nn \\ 
&&\times \int_{0}^{2\pi}\int_{0}^{2\pi}d\theta_x d\theta_y 
\log\big[\l(e^{i\theta_x}+i\r)+\l(e^{i\theta_y}-1\r)\big].\nn
\eea
In term of complex variables, i.e.  
$z=e^{i\theta_x}$ and $z'=e^{i\theta_y}$,
this can be further written into the double contour integral around a 
unit circle, 
\bea
&&\log \la 0|\eta(\vec{r})|0\ra \nn \\
&&=-N_e
\oint_{|z|=1}\frac{dz}{2\pi i}
\oint_{|z'|=1}\frac{dz'}{2\pi i} 
{\log \big[z'-z_1(z)\big]\over zz'}.
\label{z1}
\eea
In the r.h.s., we introduced $z_1(z)=-z+1-i$. 
When $|z_1(z)|>1$, the integrand of eq.~(\ref{z1}),
seen as a function of $z'$, 
has no more poles other than $z'=0$.
Thus $z'$ can be trivially integrated away.
In other words, if we decompose the contour $|z|=1$ into $C_>$ and $C_<$, where
$C_>=\{z|\ |z|=1, |z_1(z)|>1\}$ and $C_<=\{z|\ |z|=1, |z_1(z)|<1\}$,
then one can evaluate the integral along $C_>$ in the following way;
\bea
&&\int_{C_>}\frac{dz}{2\pi i}
\oint_{|z'|=1}\frac{dz'}{2\pi i} 
{\log \big[z'-z_1(z)\big]\over zz'} \nn \\
&&=\int_{C_>}\frac{dz}{2\pi i}
{\log \big[-z_1(z)\big]\over z}\nn \\
&&={3\over 8}\log 2+
{1\over \pi}\sum_{n=1}^{\infty}
\frac{\sin(n\pi/4)}{n^2 2^{n\over 2}} 
+i{9\over 16}\pi. \nn
\eea 
As for the contribution from $C_<$, one can safely verify that
it is only pure imaginary.
We also analyzed the contributions at the next leading order,
i.e. at the order of ${\cal O}(L)$, which also turns out to be pure 
imaginary (see Appendix A). We thus obtain  
the following estimation for the uniform background contributions, 
\bea
&&\log |\la 0|\eta(\vec{r})|0\ra| \nn \\
&&=-\Big[{3\over 8}\log 2+
{1\over \pi}\sum_{n=1}^{\infty}
{\sin (n\pi/4)\over 2^{n\over 2}n^2}
\Big] N_e+{\cal O}(1),
\label{eta01}
\eea
which can be evaluated as,
\be
|\la 0|\eta(\vec{r})|0\ra|=
\exp\big[-0.46 N_e+{\cal O}(1)\big].
\label{eta02}
\ee
The fact that
$\log |\la 0|\eta(\vec{r})|0\ra|< - \frac{\ln 2}{2} N_e$
in the thermodynamic limit
is one of the indispensable ingredients
for our conjecture to make sense,
the meaning of which will become clearer in the section. III.

\section{The band metal case}

Our objective here is to demonstrate that the expectation value of the 
DOP {\sl indeed vanishes for the metallic state}. 
By considering a single band metallic case, we will see explicitly 
how the presence of Fermi surface leads to the vanishing of the expectation 
value of the DOP.  
As a consequence, we are led to classify various types of Fermi surface
(or Fermi sea) into three categories by their topology. 
In the multiband case, attentions should be paid to the role
of the gauge connection in the momentum space.

\subsection{Single band --- topology of Fermi sea}
Let us first consider the simplest case, i.e., that of a single-band metal with 
$N_b=1$. This case is almost trivial in the sense
that the gauge connection  in the $\vec{k}$ space 
does not appear.
However, as we will see below, it is a good starting point
for understanding why our DOP can possibly distinguish between
a metal and an insulator.
The matrix element (\ref{zeta3ab}) reduces in the single-band case to,
\bea
&&\zeta [\vec{n}|\vec{n'}]= \sqrt{2}e^{i{3\over 4}\pi}\times \nn \\
&&\ \Big[\delta_{\vec{n},\vec{n'}}
+\Gamma_x^{(0)}\delta_{n_x,n_x'-1}\delta_{n_y,n'_y}
+\Gamma_y^{(0)}\delta_{n_x,n'_x}\delta_{n_y,n'_y-1}\Big], \nn \\
\label{zeta1b}
\eea
where
\bea
\Gamma_x^{(0)}&=&{1\over \sqrt{2}}e^{i{3\over 4}\pi}e^{-i{2\pi\over L}(x+x_0)},\nn \\
\Gamma_y^{(0)}&=&{1\over \sqrt{2}}e^{-i{3\over 4}\pi}e^{-i{2\pi\over L}(y+y_0)}. 
\label{gamma1b}
\eea
Here we have parameterized crystal momenta as 
$n_{x(y)} = 1,2,\cdots, N_{x(y)}$; 
$k_x=2\pi n_x/L, k_y=2\pi n_y/L$.  
Our Brillouin zone is thus doubly periodic, i.e., 
$n_x+N_x\equiv n_x, n_y+N_y\equiv n_y$,
forming a torus in $\vec{k}$-space. 
Although $N_x$ and $N_y$ are same, i.e. $N_x {\rm a}=N_y{\rm a}\equiv L$, 
we dare use different symbols, $N_x$ and $N_y$, only 
in this section just for clarity of the following explanations.  
The substantial simplification seen 
in eqs.~(\ref{zeta1b}) and (\ref{gamma1b}),
when compared with eq.~(\ref{zeta3ab}),
is clearly the disappearance of the gauge connections.
In the single band case, the gauge connection
associated with an overlap of two Bloch functions at different
$\vec{k}$-points gives at most a trivial phase factor, thus
erased by gauge transformations.

\subsubsection{Two representations}
In order to give an explicit matrix expression of $\zeta$
using eq.~(\ref{zeta1b}), let us introduce two representations.
The question is {\sl how to order $N_x N_y$ indices},
which we will perform in the following two steps:

\ben
\item
(${k_x,k_y}$)-representation:
The most simple and convenient way to give an explicit expression
to eq.~(\ref{zeta1b}) is to order all the $k_x$'s and $k_y$'s in the
increasing order of $n_x$ and $n_y$.
Since in our matrix representation for eq.~(\ref{zeta1b}),
each row or column corresponds to a 2D lattice point
$(n_x,n_y)$, we attribute,
(i) the inner microscopic structure (inside a given block) to 
the index $n_x$ or $k_x$, and
(ii) the outer block structure to the index $n_y$ or $k_y$.
Then, the explicit matrix element of $\zeta$ is given by
\bea
&&\zeta=\sqrt{2}e^{i{3\over 4}\pi} \times \nn \\
&&\hspace{0.3cm}\Big[ I^{(N_x N_y)}
+\Gamma^{(0)}_x \Sigma_x^{(N_x,N_y)}+\Gamma^{(0)}_y \Sigma_Y^{(N_x,N_y)}
\Big], \nn \\
\label{zeta13}
\eea
with $I^{(N)}$ being an $N\times N$ identity matrix,
\bea
\Sigma_x^{(N_x,N_y)}=\l(\ba{c|c|c|c|c}
 S_1^{(N_x)}&O^{(N_x)}&\cdots&\cdots   &O^{(N_x)} \\ \hline
O^{(N_x)}& S_1^{(N_x)}&\ddots&         &\vdots    \\ \hline
\vdots   &\ddots   &\ddots&\ddots   &\vdots    \\ \hline
\vdots   &         &\ddots& S_1^{(N_x)}&O^{(N_x)} \\ \hline
O^{(N_x)}&\cdots   &\cdots&O^{(N_x)}& S_1^{(N_x)} \ea\r), \nn
\eea
and
\bea
\Sigma_Y^{(N_x,N_y)}=\l(\ba{c|c|c|c|c|c}
O^{(N_x)}& I^{(N_x)}&O^{(N_x)}&\cdots&\cdots   &O^{(N_x)} \\ \hline
\vdots   &O^{(N_x)}& I^{(N_x)}&\ddots&         &\vdots    \\ \hline
\vdots   &         &\ddots   &\ddots&\ddots   &\vdots    \\ \hline
\vdots   &         &         &\ddots& I^{(N_x)}&O^{(N_x)} \\ \hline
O^{(N_x)}&         &         &      &O^{(N_x)}& I^{(N_x)} \\ \hline
I^{(N_x)}&O^{(N_x)}&\cdots   &\cdots&\cdots   &O^{(N_x)} \ea\r). \nn
\eea
$S_1^{(N)}$ is an $N\times N$ shift matrix defined as 
\be
S^{(N)}_1=\l(\ba{cccccc}
0     &1     &0     &\cdots&\cdots&0     \\
\vdots&0     &1     &\ddots&      &\vdots\\
\vdots&      &\ddots&\ddots&\ddots&\vdots\\
\vdots&      &      &\ddots&1     &0     \\
0     &      &      &      &0     &1     \\
1     &0     &\cdots&\cdots&\cdots&0
\ea\r). 
\ee
$\Sigma_x^{(N_x,N_y)}$ and $\Sigma_Y^{(N_x,N_y)}$ can be thus 
written symbolically as
\bea
\Sigma_x^{(N_x,N_y)}&=& S_1^{(N_x)}\bigotimes I^{(N_y)},\nn \\
\Sigma_Y^{(N_x,N_y)}&=& I^{(N_x)}\bigotimes S_1^{(N_y)}. \nn
\eea
Different subscripts, $x$ and $Y$, are used so as to recall that 
these two matrices contain a nontrivial structure 
at {\sl micro} and {\sl MACRO}-scopic levels.  

\item
Filtered-$(k_x,k_y)$-representation:
In eq.~(\ref{det2}),   
we gave an expression of the expectation value of the DOP in terms of
the determinant of $N_e\times N_e$ matrix $\zeta'$. 
This matrix, $\zeta'$, is a submatrix of $\zeta$ and composed of 
the matrix elements between {\sl occupied states}. 
In order to construct $\zeta'$, we {\sl filter} 
each state by whether it is occupied or not, 
i.e., we rearrange the order of the states 
in such a way that 
{\it occupied states are in the first $N_e$ rows and columns in 
the upper-left block of $\zeta$, 
by keeping the ordering w.r.t.~$n_x$ and $n_y$.}
We call this representation as $f$-representation in the following.
It is apparent that these two representations are related to 
each other by an orthogonal transformation. 
\een

\subsubsection{Three types of Fermi sea}
The evaluation of the determinant (\ref{det2}) 
would be trivial if all the ($N,1$)-components 
$S^{(N)}_1$ were {\sl filtered} in $\zeta'$.  
Then the matrix $\zeta'$ becomes an upper-triangular matrix and 
its determinant becomes just the product of the diagonal elements.
The question is whether this ($N,1$)-element survives
in $\zeta'$ in the $f$-representation.
This is actually dependent on the topology of Fermi sea,
which we would like to discuss now.

\begin{center}
\begin{figure}
\includegraphics[width=0.45\textwidth]{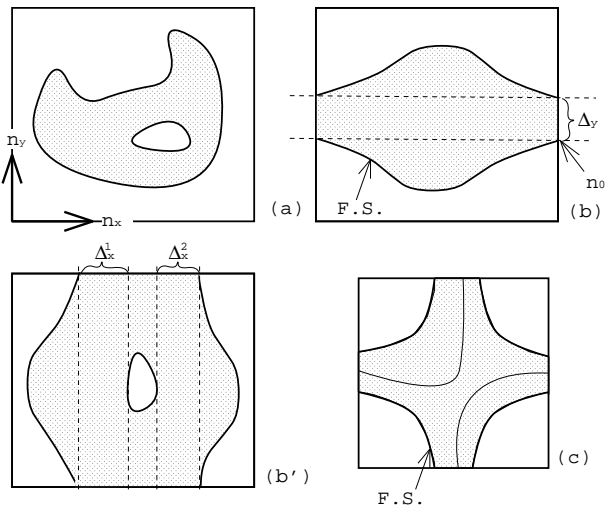}
\caption{(a) type-O Fermi surface with uneven boundary and annulus. 
(b) Type-A or B Fermi surface. The strip, which is enclosed by the two 
dashed straight line, winds the torus along $k_x$-direction.      
(b') Another type-A or B Fermi surface. 
The presence of annulus inside the Fermi sea does not change the type of 
the Fermi surface. Namely, we have only to replace 
$\Delta_y$ in eq.~(\ref{topo2}) by $\Delta^{1}_x+\Delta^{2}_{x}$. 
(c) Type-AB F.S. with a loop which winds around the torus 
along $k_{x}$ and $k_{y}$-direction.}
\label{4}
\end{figure}
\end{center}

Let us first consider a Fermi sea of trivial topology,  
which has null winding number along both 
$k_x$ and $k_y$-axes (Fig.~\ref{4}(a)).
Recall that a $(N_x,1)$-element 
of $S_1^{(N_x)}$ in $\Sigma_{x}^{(N_x,N_y)}$ is the 
matrix element between $(N_x,n_y)$ and $(1,n_y)$.  
Both of these $\vec{k}$-points are, however, 
filtered away in the $f$-representation  
{\sl for an arbitrary $n_y$}, since our Fermi sea  
does not wind the torus along the $k_x$-axis at all. 
Thus any $(N_x,1)$-elements of $S_1^{(N_x)}$ in 
$\Sigma_{x}^{(N_x,N_y)}$ do not survive in $\zeta'$. 
In a same way, one can easily see that $(N_y,1)$-{\sl block} 
of $\Sigma_{Y}^{(N_x,N_y)}$ 
is totally filtered away and does not enter into $\zeta'$ either. 
As a result, $\zeta'$ becomes {\sl an upper-triangular matrix}, 
whose determinant is identically 1 up to a trivial prefactor,
\be
\det \zeta'=2^{N_e\over 2}e^{i{3\pi\over 4}N_e}.
\ee

In order to see whether the DOP vanishes or not, we have to
compare this value with $|\la 0|\eta (\vec{r})|0\ra|$, i.e.,
eqs.~(\ref{eta01}) and (\ref{eta02}). We readily obtain,
\bea
\log|\la\eta (\vec{r})\ra|
&\simeq&\Big[
{1\over 8}\log 2-
{1\over \pi}\sum_{n=1}^{\infty}
\frac{\sin(n\pi/4)}{n^2 2^{n\over 2}}\Big] N_e
\nn \\
&\simeq&-0.11N_e.
\eea
Thus we have verified that the DOP vanishes in the thermodynamic limit
for the metallic state of trivial Fermi sea topology.

Let us now consider a less trivial case, i.e., the case of such 
a Fermi sea as depicted in Figs.~\ref{4}(b) and (b').
The Fermi sea winds the torus either along $k_x$ or $k_y$-axes.
Suppose that the Fermi sea has a filled strip 
which round the torus along the $k_x$ direction as in 
Fig.~\ref{4} (b). Then the strip is specified by 
$k_y={2\pi\over L}n_y$, where $n_y$ takes such
values as $n_y=n_0,n_0+1,\cdots,n_0+\Delta_y-1$
on the strip. The width of the strip is therefore determined by $\Delta_y$
to be $2\pi \Delta_y/L$ (see in Fig. \ref{4}(b)).

The Fermi surface topology along the $k_{y}$-axis 
is, however, still trivial. 
Thus $(N_y,1)$-block of $\Sigma_Y^{(N_x,N_y)}$ is totally 
filtered. Then the matrix $\zeta'$ turns out to be {\sl block 
upper-triangle} in the $f$-representation.  
The determinant of $\zeta'$, therefore, can be factorized into
the product of the determinants of diagonal blocks specified 
by $n_y$. 
Furthermore, for any $k_y$ or $n_y$ which is out of the strip,
the ``topology'' along the $k_x$-axis is trivial. Namely, if 
one considers a row specified by $n_y$ 
in the $(k_x,k_y)$-plane such that 
$n_y < n_0$ or $n_y > n_0 + \Delta - 1$, 
this row does not wind the torus along the $k_x$-axis.  
Thus all the submatrices (diagonal blocks) in $\zeta'$
which are specified by either  
$n_y<n_0$ or $n_y>n_0+\Delta_y-1$
are always the {\sl upper-triangular matrix},  
whose determinant is identically 1 up to a trivial 
factor.  
Then the calculation of $\det \zeta'$ reduces to those 
of $\Delta_y$ diagonal submatrices corresponding to the strip. 
They all have the form 
$\sqrt{2}e^{i{3\over 4}\pi}(I^{(N_x)}+\Gamma^{(0)}_x S^{(N_x)}_1)$. 
Since $\det [I^{(N_x)}+\Gamma^{(0)}_x S_1^{(N_x)}]$ can be readily
calculated as $1+(-1)^{N_x-1}(\Gamma^{(0)}_x)^{N_x}$,
one finally finds,  
\be
\det \zeta'=2^{N_e\over 2}e^{i{3\pi\over 4}N_e}
\Big[1+(-1)^{N_x-1} \big(\Gamma^{(0)}_x\big)^{N_x}\Big]^{\Delta_y}.
\ee
Observing
\[
1+(-1)^{N_x-1} \big(\Gamma_x\big)^{N_x}<1+2^{-{N_x\over 2}}<1+{1\over N_x}
\]
for $N_x\gg 1$, we can readily overestimate $|\det \zeta'|$ as follows, 
\be
|\det \zeta'|<2^{N_e\over 2}e^{\Delta_y\over N_x}.
\label{topo2}
\ee
In the end, we takes the thermodynamic limit, keeping the ratio
$\Delta_y/N_x$ fixed (to be smaller than 1).
Then $e^{\Delta_y\over N_x}$ gives a correction of at most
${\cal O}(1)$ in the exponential of the expectation value of the DOP, 
$|\la\eta(\vec{r})\ra| \simeq \exp \big[-0.11 N_e +{\cal O}(1)\big]$.
Thereby the correction is obviously irrelevant and the DOP 
indeed vanishes again in the thermodynamic limit.

Finally it is also possible that the Fermi sea envelopes
the torus along both $k_x$ and $k_y$-axes as 
depicted in Fig.~\ref{4}(c),
where the Fermi sea has four pockets at each corner of the
Brillouin zone (therefore still metallic).
In this case the submatrix $\zeta'$ is no longer block upper-triangle.
It is not impossible to write down formally $\det \zeta'$,
but its estimation needs some numerical analysis.
We have verified by a simple numerical analysis with substantial 
system size that expectation value of the DOP for a band metal 
having this class of Fermi surface exhibits an exponential decay  
w.r.t. the system size.

\subsection{Multiple band case without filled bands} 
When there are more than a single band, the situation becomes
suddenly complicated, because of (i) the presence of gauge connection,
and (ii) the inter-band matrix elements associated with gauge
connection. However, we can 
still give a quantitative argument by simply adding
$N_b-1$ {\sl empty} bands to the single-band case studied above.
Namely, in such a situation, we have only to calculate the determinant of
 the matrix elements in the lowest band $n=1$. In other words, 
only {\sl intra-band} ($n=1$) gauge connection appears in our analysis.
Because of this simplification, we can still use exceptionally 
the single-band equation (\ref{zeta13}), but $\Gamma_{\mu}^{(0)}$  
in eq.(\ref{zeta13}) should be replaced by the following 
$N_xN_y$ by $N_xN_y$ diagonal matrices, 
\bea
&&\Gamma_{\mu}^{(0)}\gt \nn \\
&&\ \ \ \Gamma^{(N_x,N_y)}_{\mu}[\vec{n}|\vec{n}']\equiv 
\Gamma_{\mu}^{(0)}\cdot
\la u_{1,\vec{k}}|u_{1,\vec{k}+\vec{e}_{\mu}}\ra 
\delta_{\vec{n},\vec{n}'}, \label{gamma'}
\eea
Here the unit vectors in the $\vec{k}$ space were introduced 
for convenience;   
$\vec{e}_{x}\equiv \frac{2\pi}{L}(1,0)$ and 
$\vec{e}_{y}\equiv \frac{2\pi}{L}(0,1)$. 

Let us first consider a Fermi sea of trivial topology as depicted
in Fig.~\ref{4}(a). As we have seen in the previous subsection,
the submatrix $\zeta'$, spanned by the occupied states,
becomes trivially an upper-triangular matrix in this case. Thus, 
there is clearly no room for $\Gamma^{(N_x,N_y)}_{\mu}$ to play a role.
The determinant again reduce to 1  up to a trivial prefactor
$2^{N_e\over 2}e^{i\frac{3\pi}{4}N_e}$.

The case of second topology such as Figs.~\ref{4}(b) is 
probably more interesting. 
The arguments are totally parallel until $\det\zeta'$ is 
factorized into $\Delta_y$ diagonal block matrices,
which are now not completely same.
They still have a similar form as 
$\sqrt{2}e^{i{3\over 4}\pi}(I^{(N_x)}+\Gamma^{(0)}_x S^{(N_x)}_1)$. 
But, due to the replacement of eq.(\ref{gamma'}), these 
$\Delta_y$ diagonal block matrices are also modified as,   
\bea
I^{(N_x)}+\Gamma^{(0)}_x S^{(N_x)}_1 \gt 
I^{(N_x)}+ 
\Gamma^{(N_x)}_x(n_y)\cdot S^{(N_x)}_1, \label{rep}
\eea
where $n_y$ takes such values as $n_y = n_0,n_0+1,\cdots,n_0+\Delta_y-1$ 
on the strip. The $N_x \times N_x$ diagonal matrices 
$\Gamma^{(N_x)}_{x}(n_y)$ in the r.h.s. are defined 
as $\Gamma^{(N_x)}_{x}(n_y)[n_x|{n_x}']\equiv 
\Gamma^{(N_x,N_y)}_x[n_x,n_y|{n_x}',n_y]$. 
As the number of the permutation associated with 
the determinants of both matrices in eq.~(\ref{rep})  
is same, the determinant of the r.h.s. is also readily 
calculated; 
\bea
&&\det [I^{(N_x)}+\Gamma_x^{(N_x)}(n_y)\cdot S_1^{(N_x)}] \nn \\
&&=1+(-1)^{N_x-1}\l(\Gamma_x^{(0)}\r)^{N_x}
\prod_{n_x=1}^{N_x}
\bla u_{1,\vec{k}}\big|u_{1,\vec{k}+\vec{e}_{x}}\bra. 
\eea
Hence, we have 
\bea
&&\det\zeta'=2^{N_e\over 2}e^{i{3\pi\over 4} N_e} \times \nn \\
&&\prod_{n_y=n_{0}}^{n_{0}+\Delta_y-1}
\Big\{1+ (-1)^{N_x-1} \l(\Gamma_x^{(0)}\r)^{N_x}
\prod_{n_x=1}^{N_x}
\bla u_{1,\vec{k}}\big|u_{1,\vec{k}+\vec{e}_{x}}\bra\Big\},  \nn
\eea
where $\prod_{n_x=1}^{N_x}$ is a product along
a closed loop parallel to the $k_{x}$-axis.
One might dub this loop as $C(k_{y})$ in the sense that
it is specified only by $k_{y}$. 
The product of gauge connection along this closed loop  
can be rewritten in terms of the integral of 
the abelian gauge field,  
$\la u_{1,\vec{k}}|\partial_{k_{x}} u_{1,\vec{k}}\ra$, 
w.r.t. $k_{x}$, 
\bea
&&\prod_{n_x=1}^{N_x}\bla u_{1,\vec{k}}
\big|u_{1,\vec{k}+\vec{e}_{x}}\bra \nn \\
&&=\ \exp\Big[
\int_{C(k_{y})}\la u_{1,\vec{k}}|\partial_{k_{x}} u_{1,\vec{k}}\ra\ dk_x+ 
{\cal O}\big(\frac{1}{N_x}\big)\Big]. 
\eea
Since this gauge field is pure imaginary, the amplitude of the r.h.s.  
is always unity in the thermodynamic limit. We thus conclude that 
the estimation in the previous subsection, 
i.e., eq. (\ref{topo2}),
is still unchanged even in the presence of the abelian gauge connections.
The DOP still vanishes as $\sim\exp [-0.11N_e+{\cal O}(1)]$.

In the case of the multiple band {\sl with finite number of completely filled bands}, 
we have no direct analytic proof that the expectation 
value of our DOP vanishes in metal. 
This is because all the filled bands form a torus and the 
situation associated with these filled bands is essentially same as 
the F.S. depicted in Fig.~\ref{4}(c). 
In those situations, we need to check our conjecture  
with a help of some numerics in future. Summary of this section 
is listed in the table.~II of the section.~VI.

\section{Band Insulator/Mean-field ordered state} 
Observing the results in Sec.~III that the DOP vanishes 
in the metallic case whenever it could be evaluated so far, 
we will complete our another task in this section.
Our purpose here is to verify that the expectation value of the DOP 
indeed remains finite in band insulators and gapped mean-field ordered 
states. 
In the crystal momentum space, the lowest $N_v$ bands are completely 
filled. The remaining $N_b -N_v$ bands are totally empty.
Then, on filtering away these empty bands, the expectation value of the DOP 
reads,
\[
\la\eta\ra=
\la{\rm GS}|\eta|{\rm GS}\ra=
{\det\zeta'\over (\det\zeta)^\nu}.  
\]
Here we used eq.~(\ref{remark}) as the background contributions $\la 0 | \zeta | 0\ra$.  
$\zeta'$ is the upper left part of $\zeta$ in the $f$-representation, i.e.  spanned only 
by the filled $N_v$ bands. 
Unlike the F.S. depicted in 
Figs.~\ref{4}(a), (b) and (b'), it was totally impossible, even 
in $N_v=1$ case, to factorize this determinant into the product of 
diagonal elements or diagonal blocks. Additional complications 
appear, when one considers a more general case, i.e. $1<N_v<N_b$.   
The gauge connections  between different filled bands make it 
much larger the number of the permutation associated with the 
determinant of $\zeta'$.

In order to resolve these  difficulties,
we first focus on {\sl the most strongly localized limit, i.e. the 
atomic limit}. We define this limit, such that a Hamiltonian is 
decoupled into a local Hamiltonian defined for each unit cell. 
These local Hamiltonians commute with one another. 
Thus eigenstates of a full 
Hamiltonian are given by eigenstates of each local Hamiltonian, 
which we call as {\sl localized atomic orbitals}.   
Bloch bands, which is totally flat in $\vec{k}$-space in this 
limit, can  be then obtained, on Fourier-transforming 
these localized atomic orbitals. 
The periodic part of the Bloch function $|u_{n,\vec{k}}\ra$ 
thus obtained has {\sl no $\vec{k}$-dependence} in this limit.

Starting from this atomic limit, 
we will expand $\la\eta\ra$ {\sl in powers of the number of 
$k$-derivatives of this periodic part}. This 
treatment is verified, provided that our system 
is close to the atomic limit.  
We will perform this expansion up to the second order 
and observe that $|\la\eta\ra|$ in the insulating state is 
characterized by the {\it quantum metric tensor} defined 
in eq.~(\ref{qmetric}) (See also eq.~(\ref{4-18})). 
When integrated over the crystal momentum,  
this gauge invariant 2-form is directly linked to the physical 
quantity called as the localization 
length $\xi$ (eqs.~(\ref{xiloc},\ref{global})).  
Then, we discuss that our DOP remains finite, 
as far as this localization length is finite. 
Before developing this perturbation expansion specifically,
we give a brief reminder to the relation between localization length and
metric tensor, including their physical meanings.

\subsection{brief review -- arbitrariness of Wannier function, 
its spread functional and quantum metric tensor}

The relation between 
the quantum metric tensor in the $\vec{k}$ space and localization 
length were discovered in the process of constructing 
a {\sl gauge invariant} localized Wannier function.  
Taking Fourier transformation of Bloch w.f., one can 
usually introduce a spatially ``localized'' single particle 
w.f. called  as Wannier functions,    
\bea
\la \vec{r}|\vec{R},n\ra = \frac{1}{N^2}
\sum_{\vec{k}\in {\rm Bz}}e^{i\vec{k}\cdot(\vec{r}-\vec{R})}
u_{n,\vec{k}}(\vec{r}).
\eea
However, this single particle w.f. is clearly not  invariant 
under the  $U(1)$ gauge  
transformation, $\tilde{u}_{n,\vec{k}}(\vec{r}) = e^{i\phi_{n\vec{k}}}
u_{n,\vec{k}}(\vec{r})$,
while any physical quantities should be. 
In the case of a composite set of bands, say $N_v$ filled bands, 
the corresponding gauge transformation is generalized into 
$SU(N_v)$ gauge transformations. 
In either case, shapes of Wannier functions depend   
on a particular gauge fixing and are unphysical by themselves. 

Marzari and Vanderbilt introduced a unique set of {\sl gauge independent} 
Wannier functions,  by {\sl minimizing} the spread functional 
associated with this single particle w.f.~\cite{marzari},  
\be
\Omega=\sum_{n=1}^{N_v}\Big[\la\vec{0},n|\vec{r}^2|\vec{0},n\ra
-\l(\la\vec{0},n|\vec{r}|\vec{0},n\ra\r)^2\Big],  
\ee
where measuring $\Omega$ at $\vec{R}=\vec{0}$ does not lose 
any generality. 
Namely, by optimizing the $SU(N_v)$ gauge degrees of freedom such 
that this spread functional $\Omega$ is minimized, we could 
defined the {\sl maximally localized} set of Wannier functions $\{|\vec{R}, n \ra \}$. 
This set of Wannier functions is by definition gauge invariant.   
They further prove in ref. \onlinecite{marzari} that, when optimized, 
this spread functional becomes 
\be
\Omega_I=\sum_{n=1}^{N_v}\Big[\la\vec{0},n|\vec{r}^2|\vec{0},n\ra
-\sum_{\vec{R},m}\big|\la\vec{R},m|\vec{r}|\vec{0},n\ra\big|^2\Big].
\ee
The {\it localization length} $\xi$ is introduced as this 
minimized spread; $\xi^{2}\equiv \Omega_I$.  Using the identity, 
\bea
&&\la\vec{R},n|r_\mu r_\nu \cdots|\vec{0},m\ra = \nonumber \\
&&\ \frac{1}{N^2}
\sum_{\vec{k}\in {\rm Bz}}e^{i\vec{k}\cdot\vec{R}}
\la u_{n,\vec{k}}|(i\nabla_{k_\mu})(i\nabla_{k_\nu})
\cdots|u_{m,\vec{k}}\ra, \nn 
\eea
one can then express this length $\xi$ in terms of the 
following ``metric tensor'' defined 
as~\cite{marzari,ivo,resta2},  
\bea
&&g_{\mu\nu}(\vec{k})\equiv 
\sum_{n=1}^{N_v}{\rm Re}\ \Big[ 
\bla \partial_{k_{\mu}} u_{n,\vec{k}}\big|
\partial_{k_{\nu}} u_{n,\vec{k}}\bra \nn \\ 
&&\ \ -\ \sum_{n'=1}^{N_v}
\bla\der_{k_{\mu}} u_{n,\vec{k}}\big| u_{n',\vec{k}}\bra 
\bla u_{n',\vec{k}}\big|\der_{k_{\nu}} u_{n,\vec{k}}\bra  
\Big]. \label{qmetric}
\eea
Namely, we have   
\be
\xi^2=\sum_\mu \xi_\mu^{2},\ \ \
\xi_\mu^{2}=\frac{1}{N^2}\sum_{\vec{k}\in {\rm Bz}}g_{\mu\mu}(\vec{k}).
\label{xiloc}
\ee
As is clear from this definition, $\xi$ vanishes in the atomic limit, 
where $|u_{n,\vec{k}}\ra$ has no $\vec{k}$-dependence. 

The tensor quantity $g_{\mu\nu}$ is dubbed as the {\sl quantum metric tensor}, 
in the sense that it measures the minimal {\sl distance} between 
two sets of $N_v$ eigenbasis $\{|u_{n,\vec{k}}\ra\}$ and 
$\{|u_{n,\vec{k}+d\vec{k}}\ra\}$. In fact, when optimizing the  
$SU(N_v)$ gauge degree of freedom such that  
$\sum_{n=1}^{N_v}|\la u_{n,\vec{k}}|u_{n,\vec{k}+d\vec{k}}\ra|$ is 
minimized, this {\sl quantum} distance is indeed 
specified by 
$g_{\mu\nu}(\vec{k})dk_{\mu}dk_{\nu}$~\cite{anandan}.   
 
The localization length thus introduced can  
be generalized into the following correlation function of 
{\sl macroscopic electronic polarizations}, measured w.r.t. 
the ground state many-body 
w.f. $|{\rm GS}\ra$~\cite{ivo}, 
\bea
\xi^2_{\mu}=
\frac{1}{N_e}\bla{\rm GS}\big|
\big(\hat{X}_{\mu}-\bar{X}_{\mu}\big)^2\big|{\rm GS}\bra
, \label{global}
\eea
where $\hat{X}_{\mu}\equiv\sum_{j=1}^{N_e}\hat{x}_{j,\mu}$  
denotes the total electron's position operator. 
This correlation function is infinite when a system 
has a Drude weight or peak, while remains 
finite without any Drude weight~\cite{ivo}.  
Thus, $\xi$ is also finite (infinite) in insulators (metal). 

\subsection{Disorder operator from the viewpoint of localization length}

We see below that the expectation value of 
the DOP in insulating states is simply related 
to the localization length, as given in eq.~(\ref{final}).  
To be specific, starting with the atomic limit, we expand 
$\log \la\eta\ra$ in powers of $\der_{\vec{k}}| u_{n\vec{k}}\ra$.
The zero-th order, which corresponds to the atomic limit, 
gives a constant contribution.  
At the first order, one finds a correction due to the gauge field.
However, this is pure imaginary, and irrelevant to the amplitude of
$\la\eta\ra$.
The first relevant contribution to $|\la \eta\ra|$ appears at the second order, which will be
interpreted, with the help of eqs.~(\ref{qmetric},\ref{xiloc}), as 
the square of localization length. 
Then, eq.~(\ref{final}) in combination with eq.~(\ref{global})  
tells us that the DOP remains finite in insulators, 
as far as this perturbation expansion is valid. 
Since the localization length  diverges in metals, a 
na\"ive extrapolation  of our second order 
analytic expression 
is also consistent with our previous observations found in 
the metallic case. 

\subsubsection{The zeroth order -- atomic limit}

In the atomic limit (AL), where $|u_{n,\vec{k}}\ra$ has 
no $\vec{k}$-dependence, the gauge connection becomes 
a trivial matrix, i.e. 
$\la u_{n,\vec{k}}|u_{n',\vec{k}'}\ra = \delta_{n,n'}$.  
Then the matrix $\zeta$ takes the following 
{\sl decoupled} structure in the momentum representation,
\bea
\zeta_{\rm AL}[n,\vec{k}|n^{\prime},\vec{k}^{\prime}] 
=\delta_{n,n'}\otimes \zeta_{\rm SB}[\vec{k}|\vec{k'}]. \nn
\eea
From eq.~(\ref{zeta3ab}), 
$\zeta_{\rm SB}[\vec{k}|\vec{k'}]$ is simply 
given as follows,
\bea
\zeta_{\rm SB}[\vec{k}|\vec{k'}]= 
\big(\gamma_{x}\delta_{\vec{k}+\vec{e}_x,\vec{k}'} 
+ i\delta_{\vec{k},\vec{k}'}\big) 
+ \big(\gamma_{y}\delta_{\vec{k}+\vec{e}_y,\vec{k}'} 
- \delta_{\vec{k},\vec{k}'}\big), \nn 
\eea
where $\gamma_{\mu}$ is $\sqrt{2}e^{i\frac{3\pi}{4}}\Gamma^{(0)}_{\mu}$. 
This $N^2$ by $N^2$ matrix is diagonalized by the Fourier transformation, 
where the unit cell index: $\vec{R}=(n_x {\rm a},n_y {\rm a})$, 
specifies its eigenvector, i.e. $e^{i\vec{k}\cdot\vec{R}}$. 
The corresponding eigenvalue of $\zeta_{\rm SB}$ reads,
\be
\lambda({\vec{R}})=\big(\gamma_x e^{i\vec{e}_x\cdot\vec{R}}+i\big) 
+ \big(\gamma_y e^{i\vec{e}_y\cdot\vec{R}} -1\big). 
\label{lambda}
\ee

As the band index and momentum index are 
decoupled in $\zeta_{\rm AL}$, we can easily filter unfilled bands 
so as to obtain $\log\det\zeta'_{\rm AL}={\rm Tr}\log\zeta'_{\rm AL}$. 
In other words, we have only to sum $\log\lambda(\vec{R})$ w.r.t. 
the unit cell index $\vec{R}$ and multiply the number of filled 
atomic orbital within each unit cell; 
\bea
\log\det\zeta_{\rm AL}'=N_v\sum_{n_x,n_y=1}^{N} 
\log\lambda({\vec{R}}). \nn 
\eea
Notice that $\lambda(\vec{R})$ in eq.~(\ref{lambda})  is basically  
same as the original $\zeta (\vec{R}-\vec{r})$ defined in 
eq.~(\ref{zetadef}). 
Thus we can relate the r.h.s. with $\det \zeta$ used in eq.~(\ref{remark}), 
\bea
\log\det\zeta_{\rm AL}'=\frac{N_v}{N_b}\cdot N_b \sum_{\vec{R}}\log\lambda({\vec{R}})=\nu\cdot\log\det\zeta, \nn 
\eea
where $\nu=\frac{N_e}{N_b N^2}$.
This tells us a rather trivial fact 
that our DOP becomes unit in this atomic limit, 
$\la\eta\ra_{\rm AL}=
{\det\zeta_{\rm AL}'\over (\det\zeta)^\nu}=1$~\cite{note}.  

\subsubsection{First and second orders}

We now introduce weak transfer integrals between the localized atomic 
orbitals, taking into account weak $\vec{k}$-dependences of   
$|u_{n,\vec{k}}\ra$ perturbatively. Namely, we develop 
a perturbative expansion in powers of the number of its 
$\vec{k}$-derivative. 
To concretize the perturbative evaluation 
of $\det\zeta'/(\det\zeta)^{\nu}$, 
let us consider its logarithm, 
i.e. $\log\det\zeta'={\rm Tr}\log\zeta'$,  
and decompose $\zeta'$ as $\zeta'=\zeta'_{\rm AL}+\Delta\zeta'$. 
$\Delta\zeta'$ describes the deviation from the atomic limit,
in which the information of weak delocalizations are encoded through the 
``gauge field'' $\hat{\cal A}_{\mu}(\vec{k})$, 
\bea
{\cal A}_{\mu}^{[n|n']}(\vec{k})&\equiv&
\bla u_{n,\vec{k}}\ \bigl|\ u_{n^{\prime},\vec{k}+\vec{e}_{\mu}}\bra 
-\delta_{n,n^{\prime}}\nn \\
&=&\frac{2\pi}{L}
\bla u_{n,\vec{k}}\big|\partial_{k_{\mu}} u_{n', \vec{k}}\bra  + \nn \\ 
&& \frac{1}{2}\big(\frac{2\pi}{L}\big)^2 
\bla u_{n,\vec{k}}\big|\partial^2_{k_{\mu}} u_{n', \vec{k}}\bra+
{\cal O}\big(\partial^3_{\vec{k}}\big).
\label{discrete}
\eea 
$\Delta\zeta'$ is written  in terms of this ``gauge field'' as  
\bea
\Delta\zeta'[n,\vec{k}|n^{\prime},\vec{k}^{\prime}]= 
\sum_{\mu=x,y}
\gamma_{\mu}\delta_{\vec{k}+\vec{e}_{\mu},\vec{k}^{\prime}}
{\cal A}_{\mu}^{[n|n']}(\vec{k}), \nn
\eea
where $n$ and $n'$ are restricted within filled bands.  
Observing that $\Delta\zeta'$ contains at least the first order of 
$\partial_{\vec{k}}|u_{n,\vec{k}}\ra$,  we first expand 
${\rm Tr}\log (\zeta'_{\rm AL} +\Delta\zeta')$  
w.r.t. $\Delta\zeta'$,  
\bea
&&{\rm Tr}\log(\zeta'_{\rm AL} +\Delta\zeta')=
\log{\rm det}\big[\zeta'_{\rm AL}\big] 
+{\rm Tr}\big[{\zeta'_{\rm AL}}^{-1}\Delta\zeta'\big] \nn \\
&&\hspace{1.2cm} -\  \frac{1}{2}{\rm Tr}\big[{\zeta'_{\rm AL}}^{-1}\Delta\zeta'
{\zeta'_{\rm AL}}^{-1}
\Delta\zeta'\big]+{\cal O}\big(\partial^3_{\vec{k}}\big).
\label{formalexp}
\eea

As was shown above, its zero-th order 
term is cancelled by the contribution 
from the uniform background $\la 0|\eta|0\ra=1/(\det\zeta)^\nu$, 
when exponentiated.   
On the other hand, using 
the inverse of $\zeta'_{\rm AL}$,    
\bea
{\zeta'_{\rm AL}}^{-1}[n,\vec{k}|n',\vec{k'}]=
\delta_{n,n^{\prime}}\frac{1}{N^2}\sum_{\vec{R}}
e^{i(\vec{k}-\vec{k}')\cdot\vec{R}}\lambda(\vec{R})^{-1}, \nn 
\eea 
the 1st order term w.r.t. the ``gauge field'' $\hat{\cal A}_{\mu}$  
is given as follows,    
\bea
&&{\rm Tr}\big[{\zeta'_{\rm AL}}^{-1}\Delta\zeta'\big]=
\sum_{\mu=x,y} \sum_{\vec{k}\in{\rm Bz}} \sum_{n=1}^{N_v}
{\cal A}_{\mu}^{[n|n]}(\vec{k}) \times \nn \\ 
&&\hspace{1.7cm}
{1\over N^2}\sum_{\vec{R}} \frac{\gamma_{\mu}
e^{i\vec{e}_{\mu}\cdot\vec{R}}}
{\big(\gamma_x e^{i\vec{e}_x\cdot\vec{R}}+i\big) 
+ \big(\gamma_y e^{i\vec{e}_y\cdot\vec{R}} -1\big)}. \nn
\eea
The coefficient of 
$\hat{\cal A}_{\mu}$ 
can be further evaluated by the following double 
counter integral, 
\bea
&&\frac{1}{N^2}\sum_{\vec{R}} 
\frac{\gamma_{\mu}e^{i\vec{e}_{\mu}\cdot\vec{R}}}
{(\gamma_{x} e^{i\vec{e}_{x}\cdot\vec{R}}+i)
+(\gamma_{y} e^{i\vec{e}_{y}\cdot\vec{R}}-1)} = \nn \\ 
&&\int\!\!\!\int_{[0,2\pi]\times[0,2\pi]} 
\frac{e^{i\theta_{\mu}}}{e^{i\theta_x}+ i + e^{i\theta_y}-1}
\ \frac{d\theta_x d\theta_y}{4\pi^2} 
 + {\cal O}\big(\frac{1}{L^2}\big), \nn
\eea 
where the complex variable $e^{i\theta_\mu}\equiv\gamma_{\mu}
e^{i\vec{e}_{\mu}\cdot\vec{R}}$ was introduced.  
Notice that the integrand in the r.h.s. can be 
simply replaced by  
$-i\der_{\theta_{\mu}}\{\log [e^{i\theta_x}+ i + e^{i\theta_y}-1]\}$,  
\bea
&&\int\!\!\!\int_{[0,2\pi]\times[0,2\pi]} 
\frac{e^{i\theta_{\mu}}}{e^{i\theta_x}+ i + e^{i\theta_y}-1}
\ \frac{d\theta_x d\theta_y}{4\pi^2} \nn \\
&&=-\frac{i}{4\pi^2}
\int_{0}^{2\pi}d\theta_{\bar{\mu}}
\int_{0}^{2\pi}d\theta_{\mu}\partial_{\theta_{\mu}}
\big\{\ln[e^{i\theta_x}+i+e^{i\theta_y}-1]\big\},
\nn 
\eea 
where $\bar{\mu}$ denotes the other coordinate than $\mu$ 
e.g. $\bar{x}\equiv y$. As was clear from Figs.~\ref{2},\ref{3},\ref{10}, 
$\log [e^{i\theta_x}+ i + e^{i\theta_y}-1]$ has a branch cut, 
which runs from $(-\pi/2,0)$ to $(0,-\pi/2)$ in 
the $\theta_x$-$\theta_y$ plane. 
Thus a surface term associated 
with the  $\theta_{\mu}$-integral in the r.h.s. gives 
$2\pi i$, when its integral path crosses this 
branch cut. Then, irrespective of a specific choice 
of the principle branch,  
this double counter integral is 
always estimated to be $1/4$ 
(see Fig.~\ref{10} in the case of $\mu=x$);    
\bea
&&\int\!\!\!\int_{[0,2\pi]\times[0,2\pi]} 
\frac{e^{i\theta_{\mu}}}{e^{i\theta_x}+ i + e^{i\theta_y}-1}
\ \frac{d\theta_x d\theta_y}{4\pi^2} \nn \\
&&=-\frac{i}{4\pi^2}\int_{-\frac{\pi}{2}}^{0}
d\theta_{\bar{\mu}}\cdot 2\pi i 
= \frac{1}{4}. 
\label{dd1}  
\eea  
Using this coefficient, the 1st order term 
w.r.t. $\hat{\cal A}_{\mu}$ in eq.~(\ref{formalexp}) 
then becomes as follows,   
\begin{widetext}
\bea
&&{\rm Tr}[\ \zeta^{-1}_{\rm AL}\Delta \zeta] 
= \frac{1}{4}\sum_{\mu=x,y} \int_{-\frac{\pi}{\rm a}}^{\frac{\pi}{\rm a}} 
\int_{-\frac{\pi}{\rm a}}^{\frac{\pi}{\rm a}}
dk_{x}dk_{y}\sum_{n=1}^{N_v} \Big\{
\frac{N\rm a}{2\pi}\bla u_{n}\big| \der_{k_{\mu}} u_{n}\bra 
+ \frac{1}{2}
\bla u_{n}\big|\partial^2_{k_{\mu}} u_{n} \bra \Big\} 
+ {\cal O}\Big(\frac{1}{L},\partial^3_{\vec{k}}\Big), \label{4-13}  
\eea  
\end{widetext}
where we further expanded the ``gauge field'' $\hat{\cal A}_{\mu}$ 
up to ${\cal O}(\partial^2_{\vec{k}})$, 
according to eq.~(\ref{discrete}). 

Note that the integrand in eq.(\ref{4-13}) contains the 2nd order 
derivative of $|u_{n}\ra$ w.r.t. the crystal momentum, i.e. 
$\la u_{n}|\partial^2_{k_{\mu}} u_{n}\ra$.  
As we mentioned above, this 2nd order 
derivative term is more important than the 
1st order one, i.e.  
$\la u_{n}|\partial_{k_{\mu}} u_{n}\ra$, when 
it comes to the {\sl amplitude} of $\la\eta\ra$. 
Namely, the latter one is 
clearly pure imaginary, while the real part of the 2nd order derivative 
term, i.e. $\la \partial_{k_{\mu}}u_{n}|\partial_{k_{\mu}} u_{n}\ra$,    
constitutes a metric tensor in combination  
with ${\rm Tr}[\ ({\zeta'_{\rm AL}}^{-1}\Delta \zeta')^2]$, 
which we will see below.    

${\rm Tr}[\ ({\zeta'_{\rm AL}}^{-1}\Delta \zeta')^2]$ 
has a relatively complicated form: 
\bea
&&{\rm Tr}[ {\zeta'_{\rm AL}}^{-1}
\Delta \zeta' {\zeta'_{\rm AL}}^{-1}\Delta \zeta'] = \nn  \\
&&\sum_{\mu,\nu}\sum_{\vec{k},\vec{k}^{\prime}}
\sum_{n=1}^{N_v}\sum_{n'=1}^{N_v}{\cal A}^{[n|n']}_{\mu}(\vec{k})
{\cal A}^{[n'|n]}_{\nu}(\vec{k}') \nn \\ 
&&\times \frac{1}{N^2}\sum_{\vec{R}}
\frac{\gamma_{\mu}e^{i(\vec{k}+\vec{e}_{\mu}-\vec{k}^{\prime})\cdot\vec{R}}}
{\gamma_{x} e^{i\vec{e}_{x}\cdot\vec{R}}+i 
+\gamma_{y} e^{i\vec{e}_{y}\cdot\vec{R}}-1} \nn \\
&&\times\frac{1}{N^2}\sum_{\vec{R}^{\prime}}
\frac{\gamma_{\nu} e^{i(\vec{k}^{\prime}+\vec{e}_{\nu}-\vec{k})
\cdot\vec{R}^{\prime}}} 
{\gamma_{x} e^{i\vec{e}_{x}\cdot\vec{R}^{\prime}}+i 
+\gamma_{y} e^{i\vec{e}_{y}\cdot\vec{R}^{\prime}}-1}. \label{4-14} 
\eea 
However, this can be still simplified substantially, 
by noticing that only the leading term
is relevant to our ${\cal O}(\partial^2_{\vec{k}})$-estimations, 
in the expansion of $\hat{\cal A}_{\nu}(\vec{k}^{\prime})$ around 
$\vec{k}^{\prime}=\vec{k}$, i.e. 
$\hat{\cal A}_{\nu}(\vec{k}^{\prime})= 
\hat{\cal A}_{\nu}(\vec{k})+ 
(\vec{k}^{\prime}-\vec{k})_{\lambda}\cdot \der_{k_{\lambda}} 
\hat{\cal A}_{\nu}(\vec{k})+\cdots.$  Namely, 
higher order terms in this expansion lead 
to at least the third order $\vec{k}$-derivative 
of $|u_{n,\vec{k}}\ra$,  
when substituted into Eq. (\ref{4-14}). 
Thus, we fairly replace 
$\hat{\cal A}_{\nu}(\vec{k'})$ in Eq. (\ref{4-14}) 
by $\hat{\cal A}_{\nu}(\vec{k})$.
This replacement enables us to sum over $\vec{k}^{\prime}$, 
which brings about a factor $N^2\delta_{\vec{R},\vec{R}'}$. 
Accordingly, as far as the 
${\cal O}(\partial^2_{\vec{k}})$-estimation is 
concerned, eq.~(\ref{4-14}) reduces to the following 
simple form;  
\bea
&&{\rm Tr}\big[ {\zeta'_{\rm AL}}^{-1}\Delta\zeta'
{\zeta'_{\rm AL}}^{-1}\Delta\zeta'] = \nn \\ 
&&\ \ \sum_{\mu,\nu}\sum_{\vec{k}} 
\sum_{n=1}^{N_v}\sum_{n'=1}^{N_v}
{\cal A}^{[n|n']}_{\mu}(\vec{k})
{\cal A}^{[n'|n]}_{\nu}(\vec{k}) \nn \\
&&\ \ \times \frac{1}{N^2}\sum_{\vec{R}} 
\frac{\gamma_{\mu}\gamma_{\nu}e^{i(\vec{e}_{\mu}+\vec{e}_{\nu})\cdot\vec{R}}} 
{\big(\gamma_x e^{i\vec{e}_{x}\cdot\vec{R}}+i 
+\gamma_y e^{i\vec{e}_{y}\cdot\vec{R}}-1\big)^2} 
+ {\cal O}\big(\partial^3_{\vec{k}}\big) .\nonumber 
\eea

Observing this simplification, one can readily see 
that the coefficients of $\hat{\cal A}_{\mu}\cdot\hat{\cal A}_{\nu}$ 
become diagonal w.r.t. $\mu$ and $\nu$. Evaluate  
the summation w.r.t. $\vec{R}$  in the r.h.s., in term of
the double counter integral w.r.t. $\theta_{x}$ and 
$\theta_{y}$ introduced above,  
\bea
&&\frac{1}{N^2}\sum_{\vec{R}}
\frac{\gamma_{\mu}\gamma_{\nu} 
e^{i(\vec{e}_{\mu}+\vec{e}_{\nu})\cdot\vec{R}}} 
{\big(\gamma_{x}\ e^{i\vec{e}_{x}\cdot\vec{R}}+i 
+\gamma_{y}\ e^{i\vec{e}_{y}\cdot\vec{R}}-1\big)^2} = \nn \\
&& \int\!\!\!\int_{[0,2\pi]\times[0,2\pi]} e^{i\theta_{\mu}}\ 
i\frac{\partial}{\partial \theta_{\nu}}
\big(\frac{1}  {\ e^{i\theta_x}+i 
+e^{i\theta_y}-1}\big)\ \frac{d\theta_{x}d\theta_{y}}{4\pi^2}. \nn 
\eea
Then, integrated by part w.r.t. $\theta_{\nu}$, the r.h.s. 
clearly vanishes when $\mu\ne \nu$, while  
becomes identical to 
the l.h.s. of eq.~(\ref{dd1}) when $\mu=\nu$. 
Thus, 
they are indeed diagonal w.r.t. $\mu$ and $\nu$; 
\bea
\frac{1}{N^2}\sum_{\vec{R}}\frac{\gamma_{\mu}\gamma_{\nu} 
e^{i(\vec{e}_{\mu}+\vec{e}_{\nu})\cdot\vec{R}}} 
{\big(\gamma_{x}\ e^{i\vec{e}_{x}\cdot\vec{R}}+i 
+\gamma_{y}\ e^{i\vec{e}_{y}\cdot\vec{R}}-1\big)^2} 
=\frac{1}{4}\delta_{\mu\nu} 
+ \big(\frac{1}{L^2}\big). \nn  
\eea 
Using this, the 2nd order term w.r.t. 
$\hat{\cal A}_{\mu}$ is then estimated up to 
the accuracy of ${\cal O}(\partial^2_{\vec{k}})$,  
\bea
&&{\rm Tr}\big[\ {\zeta'_{\rm AL}}^{-1}\Delta\zeta'
{\zeta'_{\rm AL}}^{-1}\Delta\zeta'] = -\ \frac{1}{4}\sum_{\mu=x,y} 
\sum_{n,n^{\prime}=1}^{N_v}
\nn \\
&&\times
\int_{\frac{-\pi}{\rm a}}^{\frac{\pi}{\rm a}}
\int_{\frac{-\pi}{\rm a}}^{\frac{\pi}{\rm a}} dk_x dk_y 
\big|\bla u_{n^{\prime}}\big|\der_{k_{\mu}} u_{n} \bra\big|^2 
+ {\cal O}\big(\frac{1}{L},\partial^3_{\vec{k}}\big). \label{4-17}
\eea

Notice the coincidence between the 
factor $\frac{1}{4}$ of eq.~(\ref{4-17}) and 
that of eq.~(\ref{4-13}). This coincidence actually helps us to 
combine these two 
contributions so as to obtain the {\sl gauge invariant}  
${\cal O}(\partial^2_{\vec{k}})$-estimation 
for the amplitude of our DOP,   
\bea
&&\ln|\la \eta(\vec{r})\ra| = -\frac{1}{8}\sum_{\mu=x,y} 
\int_{-\frac{\pi}{\rm a}}^{\frac{\pi}{\rm a}}
\int_{-\frac{\pi}{\rm a}}^{\frac{\pi}{\rm a}} dk_{x}dk_{y}
\times \nn \\
&&\sum_{n=1}^{N_v}\Big\{ 
\bla \der_{k_{\mu}} u_{n}\big|\der_{k_{\mu}} u_{n} \bra - 
\sum_{n'=1}^{N_v} 
\big|\bla u_{n^{\prime}}\big|\der_{k_{\mu}} u_{n} \bra\big|^2  
\Big\}. \label{4-18} 
\eea
Namely, the integrand is nothing but diagonal components of 
the quantum metric tensor defined in eq.~(\ref{qmetric}).  
Using the localization length 
defined in eq.~(\ref{xiloc}), 
we then rewrite this as follows,  
\bea
|\la\eta(\vec{r})\ra| = \exp\big[-{\pi^2\over 2}\bar{\xi}^2\big].
\label{final}
\eea 
where $\bar{\xi}$ is the {\sl dimensionless} localization length, 
i.e. $\bar{\xi}\equiv\frac{\xi}{\rm a}$. 
As we have already mentioned, the localization 
length is infinite (finite) when  
the system has a (no) Drude weight or peak~\cite{ivo}. 
Thus, eq.~(\ref{final}) indicates that our DOP indeed  
remains finite in band insulators/gapped mean-field 
ordered states. 

This 2nd order estimation was, however, derived perturbatively w.r.t. the 
$\vec{k}$-derivative of the periodic part of the Bloch w.f.. 
Thus eq.~(\ref{final}) obviously holds true only 
for small $\bar{\xi}$ region closed to the atomic limit.   
However, this simplest relation gives a key to extrapolate 
our analyses in this section to the metallic case studied previously. 
Namely, starting from the atomic limit where the localization length
is zero, we slowly turn on the transfer integral between 
atomic orbitals so as to increase $\bar{\xi}$. Eq.~(\ref{final}) indicates 
that increase of the localization length reduces the amplitude of 
our DOP. We could further increase the inter-atomic 
transfer such that the band gap eventually collapses and a Fermi surface 
appears. Although this situation is clearly beyond the validity of our 
expansion,  
the indication  of eq.~(\ref{final}) is still consistent with our 
analyses in Sec.~III. Namely, as the localization length diverges 
on the appearance of Fermi surface, eq.~(\ref{final}) means  
that the DOP should indeed vanish in such a metallic case.  
 
\section{``Twisting'' boundary conditions and the disorder operator} 
To help readers to obtain  
a transparent interpretation of the disorder operator, 
we will attempt in this section to clarify the similarity  
between our DOP approach and other recipes or criteria for 
probing insulating behaviors. 
A standard recipe is to investigate system's response against   
``twisting'' the boundary conditions~\cite{kohn,scalapino}. 
The twisted boundary  
condition along the $\mu$-direction requires the many-body  
wavefunction $\Phi(\vec{r}_1,\cdots,\vec{r}_{N_e})$ obeys 
\begin{eqnarray}
\Phi(\cdots,\vec{r}_i+L\  \vec{e}_{\mu},\cdots)=
e^{i\phi}\Phi(\cdots,\vec{r}_i,\cdots)
\end{eqnarray}
for arbitrary $i$, where real-valued $\phi$ can be 
non-zero modulo $2\pi$.  
With the ground state wavefunction obeying this 
twisted boundary condition, 
one can take the 2nd order derivative of its eigen-energy 
with respect to $\phi$.  
This quantity estimated at a finite system size becomes the Drude 
weight, when considered in the thermodynamic limit, i.e.   
$L, N_{e} \rightarrow \infty$.  

Notice that this twisted boundary condition can 
be transcribed into the magnetic flux $\phi$ inserted 
into those Hamiltonians {\it obeying the periodic boundary condition}~\cite{kohn,scalapino}. 
Namely, as far as its eigen-energy is concerned,  
we may as well consider the Hamiltonian defined {\it on the torus} 
having its genus pierced by the non-zero flux $\phi$.  
In the following, highlighting the {\it Aharonov-Bohm phase} 
created by the DOP, 
we will argue that applying the DOP ``successively''   
along a certain path $E$ on the 2D plane is also equivalent to 
inserting $2\pi \times {\rm (integer)}$ flux into a system;      
\be
\prod_{\vec{r}\in {\rm path}\ E} \eta(\vec{r})
\simeq e^{i\frac{2\pi}{L}\times m\sum_{\vec{r}'}r'_{\mu}\rho(\vec{r}')}.
\label{2Drel}
\ee
Here the r.h.s. is the unitary transformation, sometimes 
dubbed as ``twist operator'', which introduces $2\pi \times m$ 
flux through the genus 
associated with the $\mu$-direction. 

\subsection{flux insertion and the disorder operator}

To determine the path $E$ in eq.~(\ref{2Drel}) and 
also $\mu$ in its r.h.s., let us  
recall the detailed analytic properties 
of the DOP first,
\begin{eqnarray}
\eta(\vec{r})&=&\exp\big[
\sum_{x',y'}\log \zeta (\vec{r}'-\vec{r})
\{\rho(\vec{r'})-\bar{\rho}\}
\big] \nonumber \\
\zeta(\vec{r'}-\vec{r})&=&
-i\big(e^{i{2\pi\over L}(x'-x)}-1\big)
+\big(e^{i{2\pi\over L}(y'-y)}-1\big). \nonumber 
\end{eqnarray}
Here we have already chosen as $x_0=y_0=0$, for the sake of simplicity.
Its essential feature is actually the existence of vortex-antivortex pair.
The vortex is located at
$(x',y')=(x,y)$, whereas the antivortex at 
$(x',y')=(x+L/4,y-L/4)$.
A branch cut runs, e.g., on the straight line, $x'+y'=x+y$,
and connects the vortex with the antivortex, 
encoding the $U(1)$ phase holonomy associated with those singularities. 
In fact, one can see from Eq.~(\ref{real}) that 
${\rm Im}\log\zeta (\vec{r'}-\vec{r})$
corresponds to the ``AB phase'' which an electron at $\vec{r}'$ 
subjected under $\eta(\vec{r})$ acquires while 
traveling on the 2D plane. For example,  under the 
influence of the DOP, an electron winding 
the vortex anti-clockwise acquires $2\pi$ AB phase. 
Based on this observation, we naturally come up with the idea of representing the
analytic structure of $\eta(\vec{r})$, i.e. 
a pairwise form of the vortex and anti-vortex,  
{\sl by introducing a flux tube piercing the system
at these singularities.}

\begin{figure}
\includegraphics[width=0.4\textwidth]{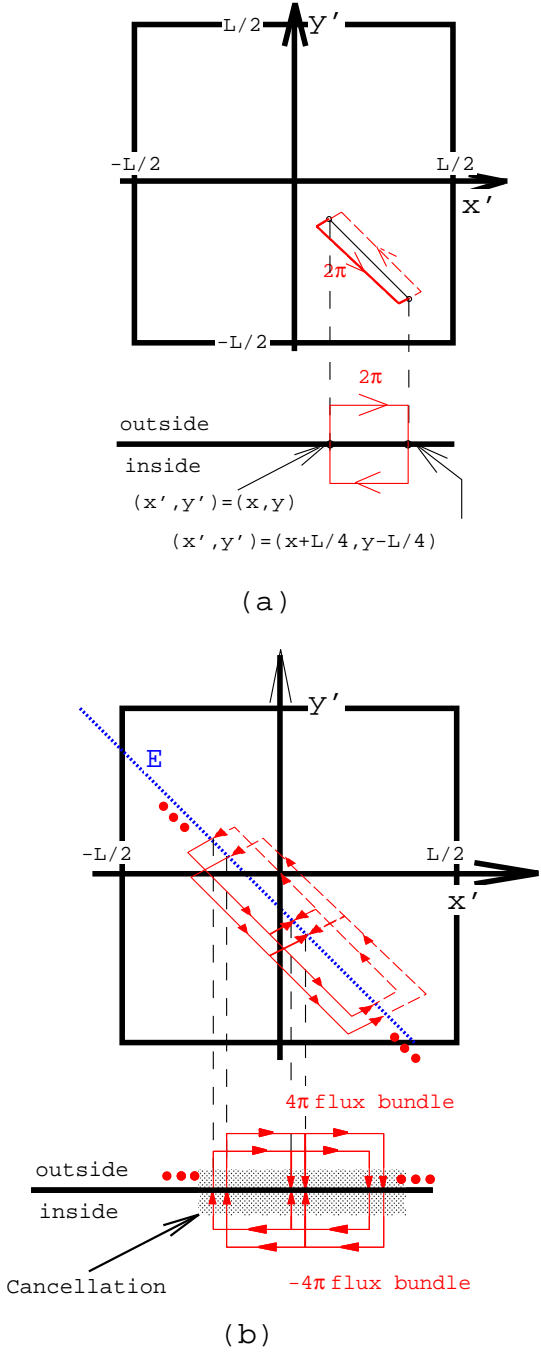}
\caption{ (Color online)  (a) A quantized ($2\pi$) flux tube associated
with the disorder operator (DOP) $\eta(\vec{r})$
is shown as a red(thin) arrows on the $(x',y')$-plane.
Corresponding to the phase inconsistency of
$\log \zeta(\vec{r'}-\vec{r})$,
the flux tube appears outside the torus at $(x,y)$,
i.e., at the vortex, and after traveling above the branch cut,
it disappears inside the torus at the antivortex, 
$(x+L/4, x-L/4)$, forming a loop.
(b) The path $E$  
is a straight blue (dotted) line along which 
a series of flux tubes associated with the DOP 
$\eta(\vec{r})$ is considered.
Flux tubes associated with
$\eta(\vec{r})$
at each lattice site $\vec{r}$ on $E$ are superposed.
In this summation over $\vec{r}$ on $E$,
cancellations  between flux tubes heading outside 
and inside the system occur on the 2D plane (shadowed region).
As a result, those flux which {\sl penetrate} 
the system completely disappear, leaving two bundles of
flux tubes above and beneath the system in parallel with $E$
(See also Fig.~\ref{6}(a)).}
\label{5}
\end{figure}

In order to materialize this point,
imagine that our 2D periodic system, forming a torus, 
resides {\sl in an 3D space}. Our flux tube also lives in this
space and intersects with the 2D plane at these two 
singular points. 
Namely, on the $(x',y')$-plane, it appears outside the torus at $(x,y)$,
i.e., at the vortex, and after traveling above our 2D plane,
it disappears inside the torus at the antivortex  
$(x+L/4, x-L/4)$, forming eventually a closed loop 
behind the system (see Fig.~\ref{5}(a). The viewpoint of 
Fig.~\ref{5} is located clearly outside the torus.)  

\begin{figure}
\includegraphics[width=0.4\textwidth]{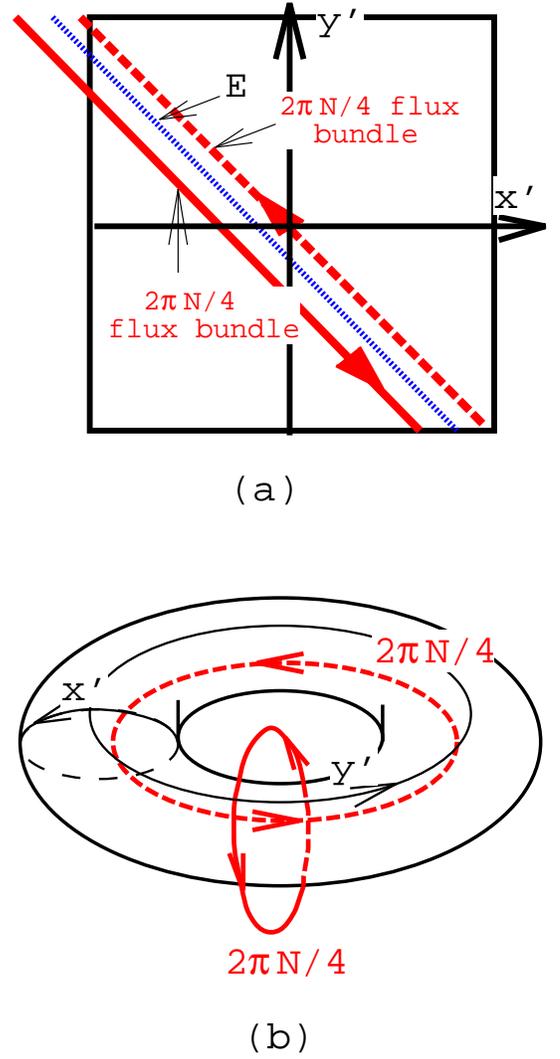}
\caption{ (Color online)
(a) Two flux bundles and the periodic 2D flat surface 
($=$ our system).
The red(solid) line represents a bundle of 
$N/4$ quantized flux tubes which flows above ($\bw$ outside the torus) 
and in parallel with our 2D plane. 
The red(broken) line is the other bundle which flows beneath ($\bw$ inside the torus)
and in parallel with the plane. 
(b) Two flux bundles and the torus ($=$ our system). 
The former pierces the handle of the torus encircled by
the $y'$-axis, whereas the latter penetrates the interior of the torus
wound by the $x'$-axis.}  
\label{6}
\end{figure}

Let us now  consider a successive application of 
$\eta(\vec{r})$ on the straight line $E$ 
parallel to $x+y={\rm constant}$, i.e.
$\prod_{\vec{r}\in E}\eta(\vec{r})$.
In the language of this flux tube, 
this corresponds to a superposition of the flux tubes
which are shifted by one lattice site
(diagonally along $E$) 
between one another (Fig.~\ref{5}(b)).
When these tubes are superposed, 
cancellations between vortices and antivortices occur 
in the vicinity of 2D plane, and {\sl the singularities on the 
plane totally disappear}. 
As a result, we are left with {\sl two decoupled vortex lines}: 
 one running above, the other beneath the system
along $E$  (Fig.~\ref{6}(a)).
These are a bundle of $N/4$ quantized ($2\pi$) flux
tubes; $N/4$, because of $N$ sites (operators) along $E$ 
and each operator having a branch cut of length, equal to
1/4 of the whole trajectory of $E$ ($= \sqrt{2} L$). 

These two vortex lines can be deformed freely 
{\sl outside or inside the torus without intersecting the torus}. 
Namely, as for an electron living in the 2D plane (torus), 
this freedom itself corresponds to a trivial $U(1)$ gauge 
degrees of freedom. Using this freedom, one can readily see that 
these two decoupled vortex lines indeed compose a pair of ``knotted'' rings,
consisting of a ``Hopf link'' (Fig.~\ref{6}(b)). In other words, we have
\be
\prod_{\vec{r}\in E}\eta(\vec{r})\simeq
e^{i\frac{2\pi}{L}\times \frac{N}{4}\sum_{\vec{r}'}(x'+y')\rho(\vec{r}')}.  
\label{zxzy}
\ee

This formula is almost correct, but there are two minor
imperfections.
First, the two sides of eq.~(\ref{zxzy}) cannot be completely 
identical, since our DOP is {\sl not} unitary, 
whereas the r.h.s.  is clearly unitary.
We, therefore, slightly soften our statement in such a way that
the {\sl unitary part} of
$\prod_{\vec{r}\in E}\eta(\vec{r})$
is equivalent to the r.h.s. in Eq.~(\ref{zxzy}). 
Secondly, since we freely deformed two vortex lines 
without intersecting our 2D plane, we can not fix  
in eq.~(\ref{zxzy}) a trivial $U(1)$ phase 
factor $e^{i\varphi_{E}}$ with a real-valued 
{\sl continuous periodic} function $\varphi_{E}(\vec{r}')$; 
\ben
\item
$\partial_{x}\partial_{y}\varphi_{E}=\partial_{y}\partial_{x}\varphi_{E}$, 
\item
$\varphi_E(x'+L,y')=\varphi_E(x',y'+L)=\varphi_E(x',y')$.
\een
Then, to the best of accuracy, our claim can be reformulated, 
in the language of the phase part of the DOP and twist operators,
as
\be
\sum_{\vec{r}}^{E}{\rm Im}\log\zeta(\vec{r}^{\prime}-\vec{r}) = 
{2\pi\over L}\times {N\over 4}(x'+y')+\varphi_E(\vec{r'}).
\label{5-6-1}
\ee

\subsection{Direct calculation of phase holonomy}

\begin{figure}
\includegraphics[width=0.45\textwidth]{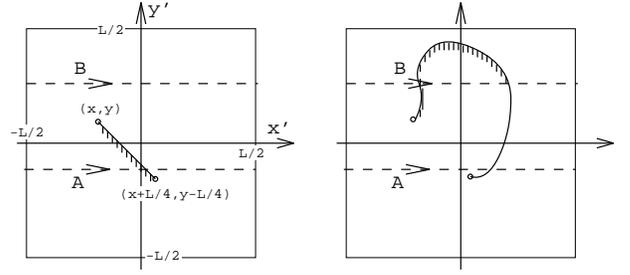}
\caption{(Left): A $x'-y'$ plane, where branch cut of ${\rm Im}\log\zeta(\vec{r}^{\prime}-\vec{r})$  
runs from $(x,y)$ to $(x+\frac{L}{4},y-\frac{L}{4})$. The shaded side of it is for 
$-\frac{\pi}{4}-$ 
while the other side is for $-\frac{9\pi}{4}+$. 
(Right): One can easily convince 
oneself that eq.~(\ref{5-10}) is independent from a specific choice of the branch cut.}  
\label{7}
\end{figure}

Eq.~(\ref{5-6-1}) elucidated by the heuristic argument 
can be directly verified, on taking lattice 
constant ``${\rm a}$'' infinitesimally small with $N{\rm a}\equiv L$ fixed.
In this continuum limit, the summation over $\vec{r}=(x,y)$ along $E$ is replaced by 
the integral w.r.t. $l=\frac{x-y}{\sqrt{2}}$ with fixed $\bar{l}\equiv \frac{x+y}{\sqrt{2}}$, 
\bea
\frac{4}{N}\sum_{\vec{r}}^{E}{\rm Im}\log\zeta(\vec{r}^{\prime}-\vec{r})
&=&\frac{4}{\sqrt{2}L}\int_{-\frac{L}{\sqrt{2}}}^{\frac{L}{\sqrt{2}}}dl 
\ {\rm Im}\log\zeta(\vec{r}^{\prime}-\vec{r})  \nn \\
&\equiv& F(x',y';\bar{l}),   
\label{5-8}
\eea 
which is a function of $\bar{l}'-\bar{l}$ in general. Then,  
by integrating $\partial_{r'_{\mu}}F$ {\sl along the $r'_{\mu}$-direction 
over  $[-\frac{L}{2},\frac{L}{2}]$}, we can 
directly calculate the AB phase which an electron acquires {\sl each time it 
travels around the periodic system along $r'_{\mu}$-direction}. Let us see 
this phase holonomy, taking $r'_{\mu}$ to be $x'$ first,   
\bea
&&\int_{-\frac{L}{2}}^{\frac{L}{2}} \frac{\partial F}{\partial x^{\prime}_{\mu}}
 dx^{\prime} = \nn \\
&&\frac{4}{\sqrt{2}L}\int_{-\frac{L}{\sqrt{2}}}^{\frac{L}{\sqrt{2}}} 
dl\ \int_{\frac{-L}{2}}^{\frac{L}{2}} 
\frac{\partial}{\partial x^{\prime}}\Big\{ {\rm Im}
\log\zeta(\vec{r}^{\prime}-\vec{r})\Big\}\ dx^{\prime}. \label{5-9}
\eea
The integral in the r.h.s. clearly vanishes due to the 
periodicity of $\log\zeta$,  in the case of  
$y^{\prime}-y\notin[-\frac{L}{4},0]$  (see the path B 
in Fig.~\ref{7}). However, as for 
$y^{\prime}-y\in [-\frac{L}{4},0]$, the path of the integral 
w.r.t. $x^{\prime}$ (the path A in Fig.~\ref{7}) 
inevitably crosses the branch cut where ${\rm Im} \log\zeta$ 
jumps by $2\pi$: ${\rm Im}\log\zeta(\vec{r}'-\vec{r})$ has 
a branch cut emitting from $\vec{r}'=(x,y)$ 
to $\vec{r}'=(x+\frac{L}{4},y-\frac{L}{4})$.  
Thus, in the latter case, the surface term gives $2\pi i$,  
on integrating w.r.t. $x'$ over 
$[-\frac{L}{2},\frac{L}{2}]$, i.e.   
\bea
&&\int_{\frac{-L}{2}}^{\frac{L}{2}} 
\frac{\partial}{\partial x^{\prime}}\Big\{ {\rm Im}\log\zeta(\vec{r}^{\prime}-\vec{r})\Big\}\ dx^{\prime} 
\nn \\
&&\hspace{0.3cm}=\ \left\{ \begin{array}{ll}
\frac{9\pi}{4}-\frac{\pi}{4} &\ \ \ \ {\rm when}\ y'-y 
\in [-\frac{L}{4},0] \\ 
0 &\ \ \ \ {\rm when}\  y'-y \notin [-\frac{L}{4},0]  \\
\end{array} \right. \nonumber 
\eea
As a result, eq.~(\ref{5-9}) reduces to  
\bea
&&\int_{-\frac{L}{2}}^{\frac{L}{2}} \frac{\partial}{\partial x^{\prime}_{\mu}}
\Big\{F(\bar{l}'-\bar{l})\Big\}\ dx^{\prime} = \nn \\ 
&&\frac{4}{\sqrt{2}L}\int_{-(\sqrt{2}y^{\prime}-\bar{l})-\frac{\sqrt{2}L}{4}}^{-(\sqrt{2}y^{\prime}-\bar{l})} dl
\ \big(\frac{9\pi}{4}-\frac{\pi}{4}\big) = 2\pi. \label{5-10}
\eea  
This unambiguously dictates 
that $F(\bar{l}'-\bar{l})$ differs from $\frac{2\pi x'}{L}$ {\sl only} by those continuous 
functions $\varphi^{\prime\prime}(x^{\prime},y';\bar{l})$ which are {\sl periodic} along the $x^{\prime}$-axis, i.e. 
$\varphi^{\prime\prime}(x^{\prime}+L,y^{\prime};\bar{l})=\varphi^{\prime\prime}(x^{\prime},y^{\prime};\bar{l})$:
\bea
F(\bar{l}'-\bar{l}) = 
\frac{2\pi x^{\prime}}{L} + \varphi^{\prime\prime}(x^{\prime},y';\bar{l}).\label{5-11}
\eea 

On repeating the same argument for $y^{\prime}$, we can easily find that these periodic functions 
$\varphi^{\prime\prime}$ is in turns different from $\frac{2\pi y^{\prime}}{L}$ only by 
those continuous functions $\varphi^{\prime}(x^{\prime},y';\bar{l})$ which 
are double-periodic {\sl both along the $x$-axis and $y$-axis}: $\varphi^{\prime}(x^{\prime}+L,y^{\prime};\bar{l})
=\varphi^{\prime}(x^{\prime},y^{\prime}+L;\bar{l})=\varphi^{\prime}(x^{\prime},y^{\prime};\bar{l})$. 
Then, we safely verified the continuum limit of eq.~(\ref{5-6-1});
\bea
&&\frac{4}{\sqrt{2}L}\int_{-\frac{L}{\sqrt{2}}}^{\frac{L}{\sqrt{2}}} dl\ 
{\rm Im}\log\zeta(\vec{r}^{\prime}-\vec{r}) \nn \\ 
&&\hspace{0.5cm} =\ \frac{2\pi (x^{\prime}+y')}{L} + \varphi^{\prime}(x^{\prime}+y';\bar{l}). 
\eea
Let us remark that the path dependence of the r.h.s., i.e. $\bar{l}\equiv \frac{x+y}{\sqrt{2}}$-dependence, 
only enters into the double-periodic function $\varphi'$ and does 
not change the strength of the flux inserted  along $x$ and $y$-direction.    
To summarize this section, we argue that the successive application of the disorder operator (DOP) 
along those arbitrary paths which parallel to its 
branch cut is equivalent to twisting 
the boundary condition of the system.

\section{Summary and Discussions}
Based on the duality picture between 2D superconductivity and insulator, 
we proposed in this paper the disorder operator as a possible candidate for
an order parameter of the 2D insulating state, which is valid for arbitrary 
2D lattice models. Namely, the change of phase of our disorder operator 
along a closed loop is nothing but the doped particle density inside this 
loop. Thus, one can naturally introduce the disorder operator (DOP) as the 
dual counterpart of the superconducting order parameter.  

To inspect the validity of this conjecture, we evaluated the expectation 
value of this non-local operator in band metal (Sec.~III) and in 
band insulators/gapped mean-field ordered state (Sec.~IV).  Thereby, 
we observed that the expectation value of the DOP actually vanishes 
for a wide class of band metals in the thermodynamic limit. 
As for band insulators, we estimated its expectation value   
perturbatively w.r.t. a weak $\vec{k}$-dependence of periodic parts 
of Bloch w.f., bearing in mind the atomic limit or weak delocalized 
insulating state close to this limit. Thereby, we found that the 
DOP is characterized by the localization length  $\bar{\xi}$ 
as $e^{-\frac{\pi^2}{2}\bar{\xi}^2}$. This localization length is 
always finite in insulators, while diverges in the presence of F.S. 
Thus, these two theoretical observations, i.e. in the metallic case 
and in the weakly delocalized insulating case, naturally lead us 
to speculate that the expectation value of the DOP in insulators 
could be always characterized by the same form, although 
our derivation in the insulator side is valid only for small $\bar{\xi}$. 
This speculation must be checked in future, with a help of 
numerics or another analytic scheme of evaluating determinants.

One might also wonder about the behaviour of the DOP in other electronic 
states, such as superconductors (SC). 
As for these off-diagonal long range order states (ODLRO), we expect   
that its expectation value should also vanish~\cite{lee1,lee2,fisher,balents1,balents2,tesanovic}, 
from the simple quantum mechanical commutation relation between a local 
density and creation operator, i.e.  
$[\rho_{\vec{r}_1},c^{\dagger}_{\vec{r}_2}] = 
\delta_{\vec{r}_1,\vec{r}_2}c^{\dagger}_{\vec{r}_2}$. 
Namely, the particle number at each 
site $\rho_{\vec{r}}$ and the phase part 
of the SC order parameter, e.g. 
$c^{\dagger}_{\vec{r},\uparrow}c^{\dagger}_{\vec{r},\downarrow}$, 
are canonical conjugate to each other.  Thus, in such an ODLRO where 
$c^{\dagger}_{\vec{r},\uparrow}c^{\dagger}_{\vec{r},\downarrow}$ 
is condensed uniformly, the particle number for each site is 
totally undermined in an entire system. 
Since the local electronic polarization 
$\vec{a}(\vec{r})$ and  phase part of the DOP, i.e. $\ln \eta(\vec{r})$,  
are given in terms of the linear combination of these local electron 
densities, they should be also {\sl indefinite}. In other word, 
the DOP should not acquire a finite {\sl amplitude} in these ODLRO states.

The standard approach of detecting insulating states (and also 
superconducting states) is to measuring the ground state energy 
variation w.r.t. the infinitesimally small Aharonov-Bohm (AB) phase 
inserted in parallel with systems~\cite{kohn,resta1,resta2,scalapino}. In order to make a 
connection to these conventional approaches, we prove that applying the DOP successively 
along a closed path also eventually ends up with the AB 
phase inserted in parallel to the 2D plane.  

\begin{widetext}

\begin{table}[htbp]
\begin{center} 
\begin{tabular}{|l|c|c|c||c|c|}
\hline
\hspace{1.0cm}ground state w.f. &\multicolumn{3}{|c||} 
{band metal}& \multicolumn{2}{|c|}{band insulator/gapped m-f state} \\ \hline
\hspace{0.0cm}type of F.S./ localization length &type.~O& type.~A or type.~B & type.~AB
& general $\bar{\xi}$ & small $\bar{\xi}$ \\ \hline \hline
\hspace{1.5cm}single band&\ $e^{-0.11\cdot N_e}$\ &\ \ $< e^{\alpha}\cdot e^{-0.11\cdot N_e}$\ 
&\ $-$\ & & \\ \cline{1-4}
\hspace{0.6cm}m-b. w/o. filled bands&\ $e^{-0.11\cdot N_e}$\ &\  \ 
$< e^{\alpha}\cdot e^{-0.11\cdot N_e}$\ &\ $-$\ & \hspace{0.9cm}  $-$ \hspace{0.9cm} 
&\ $e^{-\frac{\pi^2}{2}\bar{\xi}^2}$ \  \\ \cline{1-4}
\hspace{0.8cm}m-b. w. filled bands &\ $-$\ &\ $-$\ &\ $-$\ & &  \\ \hline
\end{tabular}
\caption{Summary of Sec.~III and IV.  ``m-b.'' denotes 
``multiple band''. Type.~O, type.~A(B) and type.~AB 
denotes a type of the F.S. 
of 2D metal (See Fig.~\ref{4}). $\bar{\xi}\equiv \frac{\xi}{\rm a}$ denotes a 
localization length along the $\mu$-direction (see eqs.~(\ref{xiloc},\ref{global}) for its definition). 
``$-$'' is the case which we cannot access by our analytical scheme.}
\end{center}
\end{table}

\end{widetext}

Observing these several circumstantial evidences including 
the connection with the conventional approach, 
we believe, in spite of its complexness   
stemming from its non-locality, that the DOP is finite 
only in an insulator. Then, turning back to our 
original motivations, i.e. {\sl microscopic identification of 
the counterpart of magnetic penetration depth and coherence 
length},  we are now allowed to push forward a naive thought on 
this primary motivations. We expect that the localization 
length $\bar{\xi}$ characterizes not only the expectation value 
of the DOP in insulating states, but it also specifies 
{\sl the counterpart of the magnetic penetration depth}, 
by the following reasons.
Notice, from eq.~(\ref{global}), that the localization 
length measures {\sl how easily the localized electrons constituting 
the background non-doped insulating w.f. could be 
polarized}, when an external electric field is applied, 
or in other words 
{\sl when a test charge is introduced into this system}. 
When this test charge is regarded as {\sl a single (few) doped particle}, 
the polarizability of the background insulating electrons then 
could be transcribed into {\sl how easily a single doped particle 
push away these background electrons  and acquires its own seat within a bulk, 
or in other words, forming a vortex within a bulk}. Thus, we  
expect that there should 
be a certain amount of positive correlation between 
the localization length $\bar{\xi}$ measured w.r.t. {\sl non-doped} gapped 
mean-field ordered w.f.  and the counterpart 
of the magnetic penetration depth in its {\sl weakly doped} 
regions. These speculation could be directly judged, when one 
construct a Ginzburg-Landau type theory for doped 2D insulating state, 
by using this non-local field, or alternatively,  when calculating 
its multi-point correlation functions, w.r.t. the 
non-doped insulating w.f..   

\acknowledgments

Authors are pleased to acknowledge S. Ryu, A. Furusaki, C. Mudry, 
Y. Hatsugai and L. Balents. R.S. 
is supported by JSPS (Japanese Society for the Promotion of Science) 
as a Postdoctoral Fellow. K.I. is supported by RIKEN 
as a Special Postdoctoral Researcher.

\appendix

\section{Contribution from the uniform background} 

\begin{figure}
\includegraphics[width=0.3\textwidth]{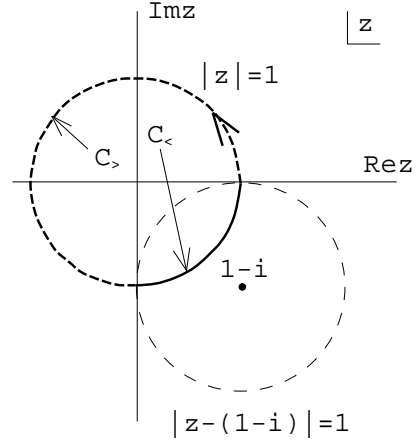}
\caption{Two integral paths w.r.t. $z$, $C_{<}$ and $C_{>}$}
\label{8}
\end{figure}

When we prove our DOP indeed vanishes in a band metal, it is crucial that 
the contribution from the uniform background  $\la 0 | \eta(\vec{r})| 0 \ra$ 
overkills the contribution from $\det\zeta'$, i.e. 
$|\la 0|\eta(\vec{r})|0\ra| \ll e^{-\frac{\ln 2}{2}N_e}$.  
Therefore, in this appendix, we will estimate the contribution from the uniform 
background,    
\bea
&&-{\rm ln}\la 0|\eta(\vec{r})|0\ra = \frac{N_e}{N_b N^{2}}\sum_{\vec{r}^{\prime}}\times \nn \\
&&\ \ {\rm ln}\big[ -i (e^{i\frac{2\pi}{N\rm a}(x^{\prime}-x-x_0)} -1) 
+ (e^{i\frac{2\pi}{N\rm a}(y^{\prime}-y-y_0)} -1)\big], \nn \\
\label{7-1}
\eea
up to the order of ${\cal O}(1)$ and prove that this is indeed the case. 
This summation can be replaced by the integral, 
as far as $O(L^{2})$ is concerned: 
\bea
&&-\ln \la 0|\eta(\vec{r})|0\ra = \frac{N_e}
{N_bN^{2}}\cdot N_b \sum_{x^{\prime},y^{\prime}={\rm a}}^{N\rm a}\times   \nn \\
&&\ \ {\rm ln}\big[ -i (e^{i\frac{2\pi}{N\rm a}(x^{\prime}-x_0)} -1) 
+ (e^{i\frac{2\pi}{N\rm a}(y^{\prime}-y_0)} -1)\big] \nonumber \\ 
&=& \frac{N_e}{4\pi^2}\int_{0}^{2\pi} d\theta_x \int_{0}^{2\pi} d\theta_y\ 
{\rm ln}\big[i + e^{i\theta_x} -1 + e^{i\theta_y}\big] \nonumber \\
&=&  N_e\ \oint_{|z|=1} \frac{dz}{2\pi i}\oint_{|z^{\prime}|=1} 
\ \frac{dz^{\prime}}{2\pi i} \ 
\frac{{\rm ln}\big[z^{\prime}-z_{1}(z)\big]}{z\cdot z^{\prime}}, \label{7-2}
\eea   
where we introduced complex variables $z\equiv e^{i\theta_x}$,  
$z^{\prime}\equiv e^{i\theta_y}$ and $z_{1}(z)\equiv -z+1-i$. 
The additional factor $N_b$ in the 1st line 
comes from the summation within the unit cell. 
Then, we will first divide the above $z$-integral into 
the following two parts, according as the position of 
$z_{1}(z)$ in the complex plane;
\bea
\oint_{|z|=1}\frac{dz}{2\pi i}=\int_{C_{>}}\frac{dz}{2\pi i}
+\int_{C_{<}}\frac{dz}{2\pi i}.  \label{7-3}
\eea
Namely, the region $C_{>,<}$ are defined as follows (see Fig.~\ref{8}):
\bea
C_{<} &\equiv& \{ z\ |\ |z|=1, \ |z_{1}(z)| < 1\}, \nonumber \\
C_{>} &\equiv& \{ z\ |\ |z|=1, \ |z_{1}(z)| > 1\}. \nonumber 
\eea 

\begin{figure}
\includegraphics[width=0.3\textwidth]{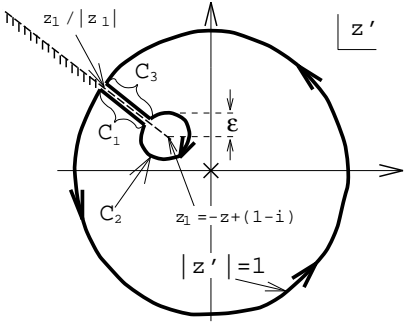}
\caption{Dotted line emitting from $z_1$ to infinity is the branch 
cut of $\ln [z'-z_1]$ which we have chosen in eq.~(\ref{7-4}). 
$\arg\ln[z'-z_1]$ takes $+0 i$ in the shaded side of this dotted line 
while $2\pi$ in the other side.}
\label{9}
\end{figure}

We can easily see that the 2nd term in eq.~(\ref{7-3}) 
does not contribute to the real part of eq.~(\ref{7-2}), 
by choosing the branch cut of the logarithm, i.e. 
$\log[z'-z_{1}]$, {\sl so that it depends on $z_1$}. 
Namely, let it run from $z_{1}$ 
to infinity {\sl passing through 
$\frac{z_{1}}{|z_{1}|}$} (see in Fig.~\ref{9}). 
Note that the real part of eq.~(\ref{7-2}) is free 
from the specific choice of the branch cut. 
According as this choice, 
we will decompose the $z^{\prime}$-integral 
into the following four parts:
\bea
\oint_{|z^{\prime}|=1} \ \frac{dz^{\prime}}{2\pi i} 
= \biggr(\oint_{C}-\int_{C_{1}}-
\oint_{C_{2}}-\int_{C_{3}}\biggl)
\cdot\frac{dz^{\prime}}{2\pi i}, \label{7-4}
\eea
where the closed path $C$ is the bold loop depicted 
in Fig.~\ref{9} and 
\bea
C_{1} &\equiv& \big\{ z^{\prime}\ |\ {\rm arg}\ln z' = 
{\rm arg}\ln z_{1} + 0,\ |z_{1}| < |z^{\prime}| < 1\big\}, \nonumber \\ 
C_{2} &\equiv& \big\{ z^{\prime}\ |\ |z^{\prime}-z_{1}|=+\epsilon\big\}, \nonumber \\
C_{3} &\equiv& \big\{ z^{\prime}\ |\ {\rm arg}\ln z' = 
{\rm arg}\ln z_{1} - 0,\ |z_{1}| < |z^{\prime}| < 1\big\}. \nonumber   
\eea  
First of all, the third term in the l.h.s. of eq.~(\ref{7-4}) 
clearly vanishes, when $\epsilon$ is taken to be 
infinitesimally small:
\bea
\oint_{C_{2}}\frac{dz^{\prime}}{2\pi i}\ 
\frac{{\rm ln}\bigl[z^{\prime}-z_1\bigr]}{z^{\prime}}
\cong \frac{i}{2\pi}\frac{\epsilon}{z_1}
\int_{0}^{2\pi}d\phi\ e^{i\phi}\phi, \label{7-5} 
\eea
where $z'-z_1 \equiv \epsilon e^{i\phi}$.  
As the integrand in eq.~(\ref{7-2}), seen as a function of $z'$, has 
no more pole than $z'=0$ inside $C$,  
the first term in eq.~(\ref{7-4}) reads:
\bea
&&\oint_{C}\frac{dz^{\prime}}{2\pi i}\ \frac{{\rm ln}
\bigr[z^{\prime}-z_1\bigl]}
{z^{\prime}} = {\rm ln}|z_1| + i\pi. \label{7-6}
\eea
Due to our choice of the branch cut, 
the second and fourth terms in eq.~(\ref{7-4}) 
can be substantially simplified:
\bea
&&\int_{C_{1}} \frac{dz^{\prime}}{2\pi i} 
\frac{{\rm ln}|z^{\prime}-z_1| + 0\cdot i }{z^{\prime}} 
\ + \ \int_{C_{3}} \frac{dz^{\prime}}{2\pi i} 
\frac{{\rm ln}|z^{\prime}-z_1| + 2\pi i}{z^{\prime}} \nonumber \\
&&\hspace{2.4cm} =\ \int_{C_3}dz^{\prime}
\frac{1}{z^{\prime}} =  {\rm ln}|z_1| \label{7-7}
\eea
Then, by combining eq.~(\ref{7-6}) and eq.~(\ref{7-7}), 
we finally obtain the 2nd term of eq.~(\ref{7-3}) 
as, 
\bea
&&\int_{C_{<}}\frac{dz}{2\pi i}\frac{1}{z}\oint_{|z^{\prime}|=1}
\frac{dz^{\prime}}{2\pi i}\frac{{\rm ln}\big[z^{\prime}-z_1\big]}
{z^{\prime}} \nn \\
&&=\frac{1}{2}\int_{C_{<}}\frac{dz}{z} 
=\frac{i}{2}\int_{-\frac{\pi}{2}}^{0} dk=\frac{\pi}{4}i. \label{7-8}
\eea
Thus, as far as the real part is concerned, there is no contribution to 
eq.~(\ref{7-2}), when $z$ is located on $C_{<}$. Then we will 
concentrate on the case where $z$ is located on $C_{>}$. 

When $z$ on $C_{>}$, $z^{\prime}$-integral can be 
easily done;
\bea
&& \int_{C_{>}}\frac{dz}{2\pi i}\frac{1}{z}
\oint_{|z^{\prime}|=1}\frac{dz^{\prime}}{2\pi i}
\frac{{\rm ln}\big[z^{\prime}-z_1(z)\big]}{z^{\prime}} \nn \\
&&\ \ = -\int_{C_{>}}\frac{dz}{2\pi i}\frac{{\rm ln}\big[-1 + i + z\big]}{z} \nonumber \\
&& \ \ = \frac{3}{8}{\rm ln}2 + \frac{1}{\pi}\sum_{n=1}^{\infty} 
\frac{1}{n^2}\big(\frac{1}{\sqrt{2}}\big)^n\sin\big(\frac{n\pi}{4}\big) 
+ i\frac{9\pi}{16} \nonumber \\ 
&& \ \ =0.464847699170805... + i\frac{9\pi}{16}. \label{7-9}
\eea 
Then, as far as the order of ${\cal O}(L^2)$ is concerned, 
logarithm of $|\la 0|\eta(\vec{r})|0\ra|$ can be estimated to be 
\bea
\ln |\la 0|\eta(\vec{r})|0\ra| = - 0.464847...\times N_e + 
{\cal O}(L). \label{7-10}
\eea 

Next, we will estimate the ${\cal O}(L)$ 
contribution to ${\rm ln}\la 0|\eta(\vec{r})|0\ra$, which 
turns out to be also pure imaginary. ${\cal O}(L)$ 
contribution comes from the error 
due to the replacement of the summation given in 
eq.~(\ref{7-1}) by the integral given in eq.~(\ref{7-2}). 
This error can be then estimated by the summation of  
the first derivative of integrand in eq.~(\ref{7-2}) 
w.r.t. $\theta_{x}$ and/or $\theta_{y}$. 
In other words, its leading order correction is 
always proportional to the following integrals 
{\sl with some real-valued coefficients}, 
\bea
\int_{0}^{2\pi}\int_{0}^{2\pi} \frac{d\theta_{x}d\theta_{y}}{4\pi^2}
\frac{\partial}{\partial \theta_{\mu}} 
\Big\{{\rm ln} \big[i + \ e^{i\theta_{x}} -1 + 
e^{i\theta_{y}}\big]\Big\}, \label{7-11}
\eea
where $\mu=x,y$. 

\begin{figure}
\includegraphics[width=0.45\textwidth]{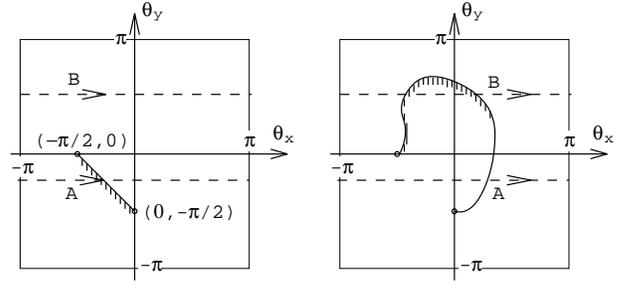}
\caption{(Left): A $\theta_{x}$-$\theta_{y}$ 
plane, where branch cut of ${\rm ln} \big[i  + \ e^{i\theta_{x}} -1 + e^{i\theta_{y}}\big]$ 
runs from $(-\frac{\pi}{2},0)$ to $(0,-\frac{\pi}{2})$. The 
shaded side of it is for $-\frac{\pi}{4}-$ while the other 
side is for $-\frac{9\pi}{4}+$. 
(Right): One can easily convince oneself that eqs.~(\ref{dd1},\ref{7-12}) are 
independent from a specific choice of the branch cut.}  
\label{10}
\end{figure}

Taking the branch cut of 
${\rm ln}\big[i + \ e^{i\theta_{x}} -1 + e^{i\theta_{y}}\big]$ 
as in Fig.~\ref{10}, 
one can easily understand these correction are pure 
imaginary. Let us see this, 
taking $\mu=x$ for example. 
In the case of $\theta_{y} \notin [-\frac{\pi}{2},0]$ (see the line B in Fig.~\ref{10}), 
the $\theta_{x}$-integral  always vanishes due to the periodicity of 
${\rm ln}\big[i + \ e^{i\theta_{x}} -1 + e^{i\theta_{y}}\big]$. On the other hand, in the case of
 $\theta_{y}\in [-\frac{\pi}{2},0]$, the integral w.r.t. $\theta_{x}$ meets $2\pi$ jump when its integral path 
crosses the branch cut which extends from 
$(-\frac{\pi}{2},0)$ to $(0,-\frac{\pi}{2})$ 
(see the line A in Fig.~\ref{10}). 
Then, when integrating w.r.t. $\theta_x$ over $[-\pi,\pi]$,  
we obtain $2\pi i$ as the surface term in the latter case.  
As a result, the above integrals with $\mu=x$ reads:
\bea
\frac{1}{4\pi^2}\int_{-\frac{\pi}{2}}^{0}
d\theta_{y} \times 2\pi i = \frac{i}{4}. \label{7-12}
\eea   
In a same way, we can easily see that the integral with $\mu=y$ also 
turns out to be $\frac{i}{4}$. 
After all, the error which stems 
from replacing the summation of eq.~(\ref{7-1}) by the integral given 
in eq.~(\ref{7-2}) can have a contribution of the order 
 ${\cal O}(L)$, but it is always 
pure imaginary.  Thus, when it comes 
to ${\rm ln}|\la 0|\eta(\vec{r})|0\ra|$, 
there is no ${\cal O}(L)$ contributions. 

To summarize this appendix, ${\rm ln}|\la 0|\eta(\vec{r})|0\ra|$ 
is estimated up to the order of ${\cal O}(1)$ and it 
reads:
\bea
\ln |\la 0|\eta(\vec{r})|0\ra| = -0.464847... \times N_e + {\cal O}(1), \label{7-13}
\eea
which is actually less than $-\frac{\log 2}{2} N_{e}$.

\bb
\bibitem{lang} 
K.M. Lang, V. Madhavan, J.E. Hoffman, E.W. Hudson, H. Eisaki, S. Uchida and J.C. Davis,  
Nature {\bf 415}, 412, (2002)
\bibitem{hana}
T.Hanaguri, C. Lupien, Y. Kohsaka, D.H. Lee, M. Azuma, M. Takano, H. Takagi, J.C. Davis, 
Nature {\bf 430}, 1001, (2004)
\bibitem{uehara}
M. Uehara, S. Mori, C.H. Chen, S-W. Cheong, 
Nature {\bf 399}, 560, (1999)
\bibitem{lee1} D.H. Lee, {\sf cond-mat/0208490} 
\bibitem{lee2}
D.H. Lee and S.A. Kivelson, 
Physical Reveiw B {\bf 67}, 024506, (2003)
\bibitem{fisher} 
M.P.A. Fisher and D.H. Lee, 
Physical Review B {\bf 39}, 2756, (1989)
\bibitem{balents1}
L. Balents, M.P.A. Fisher and C. Nayak, 
International Journal of Modern Physics B {\bf 12}, 1033, (1998) 
\bibitem{balents2}
L. Balents, M.P.A. Fisher and C. Nayak, 
Physical Review B {\bf 60}, 1654, (1999)
\bibitem{tesanovic} Z. Tesanovic,  
Physical Review Letters {\bf 93}, 217004, (2004) 
\bibitem{kohn} W. Kohn, Physical Review A {\bf 133}, A171, (1964) 
\bibitem{kudinov}
E. K. Kudinov, Soviet Physics - Solid State, {\bf 33}, 1299, (1991)
\bibitem{ivo}
I. Souza, T. Wilkens and R.M. Martin, 
Physical Review B {\bf 62}, 1666, (2000) 
\bibitem{note} Accurately speaking, 
$N_b\sum_{\vec{R}}\log\lambda(\vec{R})$ is not exactly same 
as $\sum_{\vec{r}}\log\lambda(\vec{r})$. However, as  
shown in appendix. A, their difference is at most 
order ${\cal O}(1)$, when it comes to their real part. 
Thus, the amplitude of our DOP in the atomic limit 
is also of ${\cal O}(1)$, i.e.  
\bea
\big|\eta_{\rm AL}\big|=\Big|{\det\zeta_{\rm AL}'\over (\det\zeta)^\nu}\Big|
=e^{{\cal O}(1)}. 
\eea
Furthermore, this ${\cal O}(1)$-error 
depends only on the specific choice of the position of 
$(x_0,y_0)$, introduced in eq.(\ref{zetadef}), within 
each unit cell and does not contains any information 
of the ground state w.f.. 
Thus we do not care about this contribution henceforth.  
 
\bibitem{deG} P. G. de Genne, 
{\it Superconductivity of metals and alloys} 
(Perseus Books, Massachusetts, 1999).  
\bibitem{frad} E. Fradkin and L. P. Kadanoff,
Nuclear Physics {\bf B 170}, 1 (1980).
\bibitem{kivelson}
S. Kivelson,
Physical Review B {\bf 26}, 4269, (1982)
\bibitem{marzari}
N. Marzari and D. Vanderbilt,  
Physical Review B {\bf 56}, 12847, (1997)
\bibitem{resta1} 
R. Resta  
Physical Review letters {\bf 80}, 1800 (1998). 
\bibitem{resta2} 
R. Resta and S. Sorella, 
Physical Review Letters {\bf 82}, 370, (1999). 
\bibitem{scalapino}
D.J. Scalapino, S.R. White, S.C. Zhang, 
Physical Review B {\bf 47}, 7995 (1993); 
Physical Review Letters {\bf 68}, 2830 (1992).  
\bibitem{anandan} 
J. Anandan and Y. Aharonov, Physical Review Letters {\bf 65}, 
1697 (1990).
\eb 
\end{document}